\documentclass[twocolumn]{aastex631}
\usepackage{CJK}
\usepackage{amsmath}
\DeclareMathOperator{\arctantwo}{arctan2}
\shorttitle{Tomographic Imaging of Sagittarius Spiral Arm's Magnetic Fields}
\shortauthors{Doi et al.}
\graphicspath{{./}{figures/}}
\begin{document}    
\begin{CJK*}{UTF8}{ipxm}%{gbsn}
\title{Tomographic Imaging of the Sagittarius Spiral Arm's Magnetic Field Structure}

\correspondingauthor{Yasuo Doi}
\email{doi@ea.c.u-tokyo.ac.jp}

\author[0000-0001-8746-6548]{Yasuo~Doi (土井靖生)}
\affiliation{Department of Earth Science and Astronomy, Graduate School of Arts and Sciences, The University of Tokyo, 3-8-1 Komaba, Meguro, Tokyo 153-8902, Japan}

\author{Kengo~Nakamura (中村謙吾)}
\affiliation{Department of Physics, Hiroshima University, 1-3-1 Kagamiyama, Higashi-Hiroshima, Hiroshima 739-8526, Japan}

\author[0000-0001-6099-9539]{Koji S. Kawabata (川端弘治)}
\affiliation{Hiroshima Astrophysical Science Center, Hiroshima University, 1-3-1 Kagamiyama, Higashi-Hiroshima, Hiroshima 739-8526, Japan}
\affiliation{Department of Physics, Hiroshima University, 1-3-1 Kagamiyama, Higashi-Hiroshima, Hiroshima 739-8526, Japan}
\affiliation{Core Research for Energetic Universe (CORE-U), Hiroshima University, 1-3-1 Kagamiyama, Higashi-Hiroshima, Hiroshima 739-8526, Japan}

\author[0000-0002-6906-0103]{Masafumi Matsumura (松村雅文)}
\affiliation{Faculty of Education  \& Center for Educational Development and Support, Kagawa University, 1-1 Saiwai-cho, Takamatsu, Kagawa 760-8522, Japan}

\author[0000-0001-6156-238X]{Hiroshi Akitaya (秋田谷洋)}
\affiliation{Planetary Exploration Research Center, Chiba Institute of Technology, 2-17-1 Tsudanuma, Narashino, Chiba 275-0016, Japan}
\affiliation{Hiroshima Astrophysical Science Center, Hiroshima University, 1-3-1 Kagamiyama, Higashi-Hiroshima, Hiroshima 739-8526, Japan}

\author[0000-0002-0859-0805]{Simon Coud\'e}
\affiliation{Department of Earth, Environment and Physics, Worcester State University, Worcester, MA 01602, USA}
\affiliation{Center for Astrophysics \textbar\ Harvard \& Smithsonian, 60 Garden Street, Cambridge, MA 02138, USA}

\author[0000-0002-9459-043X]{Claudia V. Rodrigues}
\affiliation{Divis\~{a}o de Astrof\'{i}sica, Instituto Nacional de Pesquisas Espaciais (INPE/MCTI), Av. dos Astronautas, 1758, S\~{a}o Jos\'{e} dos Campos, SP, Brazil}

\author[0000-0003-2815-7774]{Jungmi Kwon (權靜美)}
\affiliation{Department of Astronomy, Graduate School of Science, The University of Tokyo, 7-3-1 Hongo, Bunkyo-ku, Tokyo 113-0033, Japan}

\author[0000-0002-6510-0681]{Motohide Tamura (田村元秀)}
\affiliation{Department of Astronomy, Graduate School of Science, The University of Tokyo, 7-3-1 Hongo, Bunkyo-ku, Tokyo 113-0033, Japan}
\affiliation{Astrobiology Center, Osawa, Mitaka, Tokyo 181-8588, Japan}
\affiliation{National Astronomical Observatory of Japan, Osawa, Mitaka, Tokyo 181-8588, Japan}

\author[0000-0001-8749-1436]{Mehrnoosh Tahani}
\affiliation{Banting and KIPAC Fellowships: Kavli Institute for Particle Astrophysics \& Cosmology (KIPAC), Stanford University, Stanford, CA 94305, USA}

\author[0000-0002-1580-0583]{Antonio Mario Magalh\~{a}es}
\affiliation{Instituto de Astronomia, Geof\'{i}sica e Ci\^{e}ncias Atmosf\'{e}ricas, Universidade de S\~{a}o Paulo (IAG-USP), R. do Mat\~{a}o, 1226, S\~{a}o Paulo, SP 05508-090, Brazil}

\author[0000-0001-6880-4468]{Reinaldo Santos-Lima}
\affiliation{Instituto de Astronomia, Geof\'{i}sica e Ci\^{e}ncias Atmosf\'{e}ricas, Universidade de S\~{a}o Paulo (IAG-USP), R. do Mat\~{a}o, 1226, S\~{a}o Paulo, SP 05508-090, Brazil}

\author[0000-0001-5016-5645]{Yenifer Angarita}
\affiliation{Department of Astrophysics/IMAPP, Radboud University, PO Box 9010, 6500 GL Nijmegen, The Netherlands}

\author[0000-0003-0400-8846]{Jos\'{e} Versteeg}
\affiliation{Department of Astrophysics/IMAPP, Radboud University, PO Box 9010, 6500 GL Nijmegen, The Netherlands}

\author[0000-0002-5288-312X]{Marijke Haverkorn}
\affiliation{Department of Astrophysics/IMAPP, Radboud University, PO Box 9010, 6500 GL Nijmegen, The Netherlands}

\author[0000-0003-1853-0184]{Tetsuo Hasegawa (長谷川哲夫)}
\affiliation{National Astronomical Observatory of Japan, Osawa, Mitaka, Tokyo 181-8588, Japan}

\author[0000-0001-7474-6874]{Sarah Sadavoy}
\affiliation{Department for Physics, Engineering Physics and Astrophysics, Queen's University, Kingston, ON K7L 3N6, Canada}

\author{Doris Arzoumanian}
\affiliation{National Astronomical Observatory of Japan, Osawa, Mitaka, Tokyo 181-8588, Japan}

\author[0000-0002-0794-3859]{Pierre Bastien}
\affiliation{Institut de Recherche sur les Exoplan\`etes (iREx), Universit\'e de Montr\'eal, D\'epartement de Physique, C.P. 6128 Succ. Centre-ville, Montr\'eal, QC H3C 3J7, Canada}
\affiliation{Centre de Recherche en Astrophysique du Qu\'ebec (CRAQ), Universit\'e de Montr\'eal, D\'epartement de Physique, C.P. 6128 Succ. Centre-ville, Montr\'eal, QC H3C 3J7, Canada}



\begin{abstract}
The Galactic global magnetic field is thought to play a vital role in shaping Galactic structures such as spiral arms and giant molecular clouds. However, our knowledge of magnetic field structures in the Galactic plane at different distances is limited, as measurements used to map the magnetic field are the integrated effect along the line of sight.
In this study, we present the first-ever tomographic imaging of magnetic field structures in a Galactic spiral arm.
Using optical stellar polarimetry over a $17\arcmin \times 10\arcmin$ field of view, we probe the Sagittarius spiral arm.
Combining these data with stellar distances from the {\it Gaia} mission, we can isolate the contributions of five individual clouds along the line of sight by analyzing the polarimetry data as a function of distance. 
The observed clouds include a foreground cloud ($d < 200$ pc) and four clouds in the Sagittarius arm at 1.23 kpc, 1.47 kpc, 1.63 kpc, and 2.23 kpc. The column densities of these clouds range from 0.5 to $2.8 \times 10^{21}~\mathrm{cm}^{-2}$.
The magnetic fields associated with each cloud show smooth spatial distributions within their observed regions on scales smaller than 10 pc and display distinct orientations. The position angles projected on the plane-of-sky, measured from the Galactic north to east, for the clouds in increasing order of distance are $135^\circ$, $46^\circ$, $58^\circ$, $150^\circ$, and $40^\circ$, with uncertainties of a few degrees. Notably, these position angles deviate significantly from the direction parallel to the Galactic plane.
\end{abstract}



\section{Introduction} \label{sec:intro}

Magnetic fields significantly contribute to the hydrostatic balance in the interstellar medium \citep{1990ApJ...365..544B,2001RvMP...73.1031F,2005ARA&A..43..337C,2017ARA&A..55..111H}.
Magnetic pressure and magnetic tension caused by magnetic fields are both non-uniform forces acting perpendicular to the magnetic field lines.
Therefore, magnetic fields are believed to introduce anisotropy in the gas motion and consequently have a significant impact on structure formation and evolution in the interstellar medium (ISM), ranging from galaxy formation to the formation of filamentary molecular clouds within a single star-forming region \citep{2005LNP...664..137H,2018JCAP...08..049B}.
Indeed, magnetic field lines are expected to be influenced by the motion of the interstellar medium, leading to their dragging or bending \citep[e.g.,][]{2021ApJ...923L...9D,2022FrASS...9.0027T,2023ApJ...944..139T}.
As a result, the interstellar magnetic field structure is expected to be inscribed with a history of deformation of the ISM \citep{2018MNRAS.480.2939G}.
In other words, by revealing the structure of the interstellar magnetic field, we can elucidate the formation history of the ISM structure \citep[e.g.,][]{2022A&A...660L...7T,2022A&A...660A..97T,2022FrASS...9.0027T}.
Mapping the distribution of magnetic fields from the spatial scale of individual molecular clouds to Galactic scales (10 pc -- 1 kpc scale) may therefore provide critical information for understanding the role of, for example, Galactic spiral arms in the formation of giant molecular clouds and the subsequent star formation inside them \citep[e.g.,][]{2017ARA&A..55..111H,2018ApJ...864..153Z,2022ApJ...926L...6S}.

The structure of magnetic fields can be studied by observing polarized radiation arriving from astronomical objects.
Asymmetric dust particles irradiated by incoming radiation fields align their rotation axes parallel to the ambient magnetic field direction (radiative alignment torques; \citealp{2007MNRAS.378..910L}).
This process causes polarized light from both extincted background stars and thermal dust emission from the grains themselves \citep{1966ApJ...144..318S,1988QJRAS..29..327H}.
Thus, the plane-of-sky (POS) component of the magnetic field ($B_\mathrm{POS}$), associated with dust particles that are primarily in cold neutral ISM \citep[$\lesssim 100$ K;][]{1995ASPC...80..292M}, can be observed with both stellar optical/near-infrared polarimetry and sub-mm polarimetry \citep{2007JQSRT.106..225L}.
However, one of the limitations of these observational techniques, particularly polarimetry of optically thin dust emission, is that it can only obtain the average value of the superimposed magnetic field components along the line of sight (LOS).
Especially for regions close to the Galactic plane, multiple clouds can be along the LOS and that can complicate the inferred $B_\mathrm{POS}$ from optically thin dust.

In recent years, {\it Gaia} data have provided accurate distances to stars \citep{2016A&A...595A...1G,2022arXiv220800211G,2021AJ....161..147B} and interstellar extinction values for these stars \citep{2022arXiv220606138A,2022arXiv220605989B}.
By combining these pieces of information with stellar polarimetry data, it becomes possible to reveal the three-dimensional (3D) distribution of the interstellar medium (ISM) and its associated magnetic field up to distances of a few kpc \citep[e.g.,][]{2019ApJ...872...56P,2021ApJ...914..122D,2023A&A...670A.164P}.

The Galactic magnetic field is expected to be nearly parallel to the Galactic disk, i.e., $B_\mathrm{Z} \simeq 0$, and correlated with the spiral arms \citep{2013lsmf.book..215B,2013pss5.book..641B,2015A&ARv..24....4B,2015ASSL..407..483H,2017ARA&A..55..111H}.
Polarimetry of dust emission shows a magnetic field distribution that is generally parallel to the Galactic plane \citep{2003ApJ...583L..83N,2006ApJ...648..340L,2011ApJ...741...81B,2013ApJS..208...20B,2016A&A...594A...1P}.
On the other hand, the magnetic field of the neutral ISM traced by stellar polarimetry is not always parallel to the Galactic plane \citep{2000AJ....119..923H,2020ApJS..249...23C,2022RAA....22g5003C}, and a variation of the position angle along the LOS has been observed \citep{2014AJ....148...49P,2020PASJ...72...27Z}.
We need more detailed observational information to reveal the magnetic field structure along the LOS \citep[e.g.,][]{2019Galax...7...52J}.

The Sagittarius arm is one of the four major spiral arms of the Galaxy and is observed in $-14^\circ \lesssim l \lesssim +50^\circ$ of the Galactic plane \citep{2022NewA...9701896V}.
This structure is the closest major spiral arm in the inner Galactic plane and harbors massive star-forming regions such as M8, M16, M17, and M20 \citep{2021A&A...651L..10K}.
The $l \gtrsim +20^\circ$ region is heavily obscured by the Aquila Rift in the foreground (approximately 200 to 500 pc along the LOS), but there are no noticeable foreground clouds in $l \lesssim +20^\circ$.
In addition, the arm is almost entirely in the POS in the smaller Galactic longitude range ($l \lesssim +20^\circ$), allowing us to estimate the large-scale magnetic field structure that follows the Galactic arm structure with good approximation from the observed position angle of $B_\mathrm{POS}$.
Furthermore, target stars are more abundant in the inner Galactic plane than in the outer Galactic plane, making it a good target for obtaining the 3D magnetic field from stellar polarimetry.

To reveal the magnetic field structure along the LOS in the Sagittarius arm by a stellar polarimetric survey, this paper, as a first step, will demonstrate that we can identify multiple ISM clouds and their associated local magnetic field structure along the LOS, including the amplitude of the direction dispersion of the turbulent magnetic fields.

This paper is organized as follows.
In Section \ref{sec:obs_reduct}, we describe the selection of the observation area within the Sagittarius arm, the observations, and the data reduction procedure.
Section \ref{sec:results} provides a detailed analysis of the distance dependence of the observed magnetic field position angles along the LOS. It discusses the identification of clouds through statistical analysis of the polarimetry data, as well as the magnetic field characteristics specific to each cloud.
Section \ref{sec:discussion} discusses the relationship between the observed distance dependence and the magnetic field traced by sub-mm polarimetry observed by the Planck satellite, which is integrated along the LOS, as well as the amplitude of the turbulent magnetic field in each cloud.
In Section \ref{sec:summary}, we summarize the results.

\section{Observations \& Data Reduction} \label{sec:obs_reduct}

\subsection{Target Selection}

We selected the Sagittarius arm, the nearest major spiral arm in the inner Galactic plane with abundant observable stars in optical polarimetry, as our first target to create a tomographic image of the magnetic field in a spiral arm.
We observed a target field between $+10^\circ < l < +20^\circ$ to avoid the Aquila Rift and to have a good sky position from the Higashi-Hiroshima Observatory (see Section \ref{sec:observation}).

To define the target region, in addition to the above constraints, we also imposed the following conditions,
referring to the {\it Gaia} Data Release 2 \citep[DR2,][]{2018A&A...616A...1G} catalog, which was the latest {\it Gaia} release when the observation was planned:
\begin{enumerate}
\item A sufficient number of stars ($\gtrsim 100$ stars) with {\it Gaia} distances are distributed across all distances up to $\sim 3$ kpc;
\item The interstellar extinction increases gradually with distance along the LOS, rather than experiencing concentrated increases at specific distances.
\end{enumerate}
These conditions impose a continuous sampling of the magnetic field across the Sagittarius arm along the LOS.
Consequently, we selected a $17\arcmin \times 10\arcmin$ field centered at $l=+14^\circ\negthinspace.15,~b=-1^\circ\negthinspace.47$.

\subsection{Observations}
\label{sec:observation}

We obtained linear polarimetry in the Cousins $R$ band ($R_\mathrm{C}$-band: $\lambda = 0.65~\mu\mathrm{m}$) using the Hiroshima Optical and Near-InfraRed camera \citep[HONIR;][]{2014SPIE.9147E..4OA} on the 1.5-m Kanata Telescope, Higashi-Hiroshima Observatory, on August 5, 2021.
The optics of the HONIR instrument consists of a rotating half-wave plate, a focal mask of five equally spaced slits with a 50\% opening ratio, and a Wollaston prism that splits the incident light into two orthogonally polarized images next to each other on the detector focal plane \citep[see Section 5 of][]{2014SPIE.9147E..4OA}.
As a result, five pairs of images with orthogonal polarizations are exposed across the entire surface of the detector.
To cover the $7\arcmin\negthinspace.0 \times 9\arcmin\negthinspace.6$ detector's field-of-view (FOV) with multiple exposures, we made $3 \times 3$ spatial dithers with a $31\arcsec\negthinspace.2$-step in the East-West direction and a $20\arcsec\negthinspace.0$-step in the North-South direction.

To measure the polarization parameters $q \equiv Q/I$ and $u \equiv U/I$ of each star, we acquired photometry with four position angles of the half-wave plate at $0^\circ$, $45^\circ$, $22^\circ\negthinspace.5$, and $67^\circ\negthinspace.5$ \citep{1999PASP..111..898K}.
As a result, we obtained a total of 36 exposures, with each exposure lasting 75 seconds.

We covered the target field using two adjacent FOVs centered at $l=+14^\circ\negthinspace.11,~b=-1^\circ\negthinspace.41$ and $l=+14^\circ\negthinspace.18,~b=-1^\circ\negthinspace.53$. The combined FOV size was $17\arcmin\negthinspace.0 \times 10\arcmin\negthinspace.5$ as shown in Figure \ref{fig:Sagittarius_PAgal}.

We measured stellar intensities by aperture photometry using \textit{SEXtractor} \citep{1996A&AS..117..393B}.
The typical size of the point spread function is $\sim 1\arcsec\negthinspace.8$, and we fixed an aperture diameter of $6\arcsec\negthinspace.5$ (24 pixels).

\subsection{Calibration}
\label{sec:calibration}

We calibrated the instrumental polarization by observing the unpolarized standard star G 191-B2B on July 27, 2021.
The measured instrumental polarization, an offset vector to the origin in the $q\mathchar`-u$ parameter space, is $q_\mathrm{inst} = 0.01 \pm 0.02$\% and $u_\mathrm{inst} = -0.04 \pm 0.02$\%, which is negligible for our measurements.
The stability of the instrumental polarization, measured over a period of 10 months including the observational period, is consistently better than 0.1\%, and is thus considered negligible for our measurements.
The variation of the instrumental polarization across the detector is better than 0.1\% and can also be considered negligible \citep{2014SPIE.9147E..4OA}.

We calibrated the polarization position angle by observing the strongly polarized standard stars BD+64 106, BD+59 389, and HD 204827 \citep{1992AJ....104.1563S} on July 27 and August 30, 2021.
The achieved calibration accuracy is better than $0^\circ\negthinspace.4$ and the stability during the observational period is estimated to be better than $0^\circ\negthinspace.3$.

We calibrated the polarization efficiency of the instrument by observing an artificially polarized star through a wire-grid polarizer inserted before the half-wave plate.
The measured efficiency is $99.1\pm 0.01$ \%, by which we scaled the observed polarization fractions.

We converted the measured normalized Stokes parameters, $q$ and $u$, defined in equatorial coordinates, into the Galactic coordinates, $q_\mathrm{Gal}$ and $u_\mathrm{Gal}$.
This transformation allows us to align the polarization measurements with the Galactic coordinate system for further analysis and interpretation.
The details of the coordinate conversion process are described in Appendix \ref{sec:coord_conv}.

\subsection{{\it Gaia} Identification and Selection}
\label{sec:GaiaID}

We referred to the {\it Gaia} Data Release 3 \cite[DR3,][]{2022arXiv220800211G} catalog and cross-match the observed stars with detections of polarization within a search radius of $1 \arcsec$.
We referred to a {\it Gaia}-based catalog by \cite{2021AJ....161..147B} for the distances of each star.
Among their distance estimations, we adopted `geometric' distances, including distance estimates for all our observed stars.

We limited our search by applying the condition that the renormalized unit weight error (`ruwe') $\leqslant 1.4$ and `parallax\_over\_error' $\geqslant 3$ in the {\it Gaia} DR3 catalog, and distance uncertainty (a 68\% confidence interval) $\leqslant$ 20\% of the stellar distance.
In addition, we selected data with an estimated error $\delta P \leqslant 0.3\%$ for the fractional polarization, which was typically achieved by the stars with $R_\mathrm{C} \leqslant 15.5$ mag.
Following this procedure, we identified 184 stars within the observed field.
In investigating the interstellar extinction in the observed region, we referred to 259 stars meeting the criteria of distance uncertainty $\leqslant 20\%$ and $A_\mathrm{G}$ values available in the DR3 catalog. There were 130 stars found in both datasets. We analyzed all available data for both polarization and extinction, regardless of their availability in the other dataset.
We summarize the identified stars in Table \ref{tab:data_list} in Appendix \ref{sec:data_list}.

\section{Results} \label{sec:results}

\subsection{Spatial and Distance Distribution of Polarimetry Data}

\begin{figure}[tp]
\centering
\includegraphics[width=\linewidth]{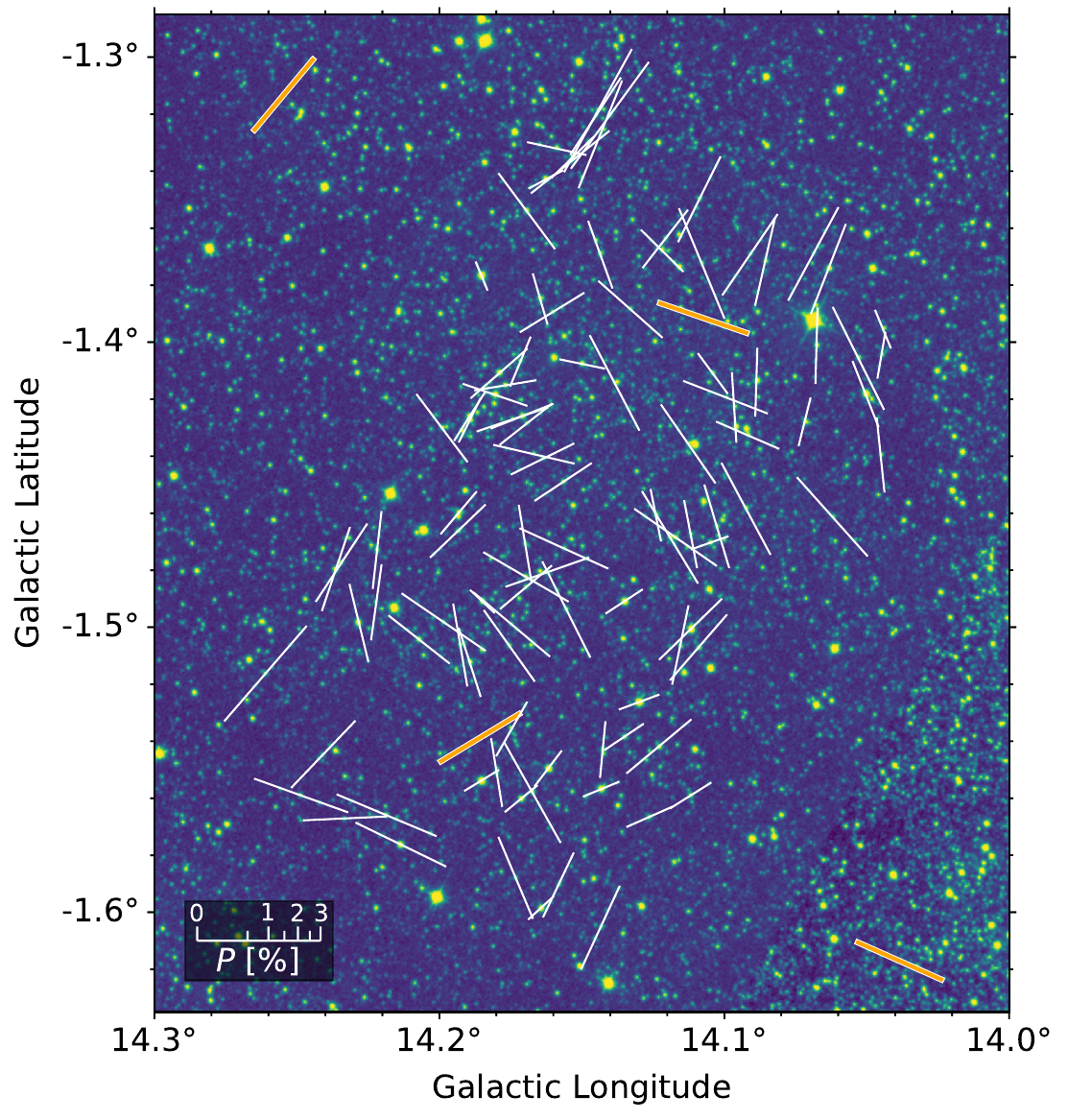}
\caption{
Observed stellar polarization pseudo-vectors (white line segments).
The data of 105 stars with errors on $P\!A$ below $\delta P\!A \leqslant 10^\circ$ are shown, out of 184 stars with significant polarization detection and accurate distance estimation.
A reference scale of $P$ is shown in the lower left corner of the figure.
The background is from the Second Generation Digitized Sky Survey red image \citep{2000ASPC..216..145M}.
The orange line segments are magnetic field position angles obtained from Planck data at 353 GHz \citep[][resolution set to $10\arcmin$]{2020A&A...641A...1P}.
The Planck's line segments only show the orientation of the magnetic field, estimated by rotating the polarization $P\!A$ by $90^\circ$, and their length is not related to the Planck-measured polarization degree.
}
\label{fig:Sagittarius_PAgal}
\end{figure}

Figure \ref{fig:Sagittarius_PAgal} shows the spatial distribution of the observed polarization pseudo-vectors (white segments), indicating position angles ({\it P\!A}) and polarization fraction ({\it P}). The derivation of these values from the observed $q$ and $u$ values is detailed in Appendices \ref{sec:coord_conv} and \ref{sec:data_list}.
Of the 184 stars applied in the following analyses, 105 stars with a polarization position angle uncertainty $\delta P\!A \leqslant 10^\circ$ are plotted in the figure.
The distribution of $B_\mathrm{POS}$ traced by stellar polarimetry appears to be a perfect mix of various {\it P\!A}s in space.
\begin{figure}[tp]
\centering
\includegraphics[width=\linewidth]{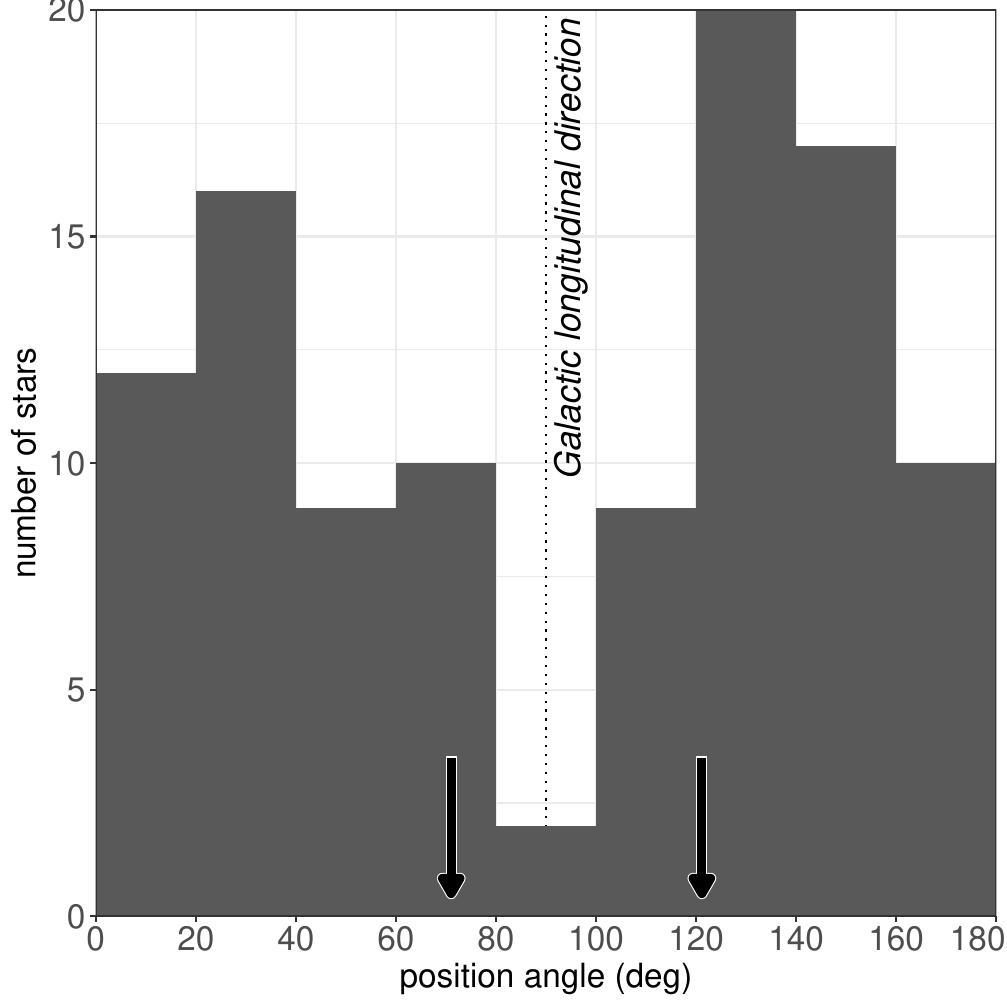}
\caption{
Histogram of $P\!A$.
The bin width is set as $20^\circ$.
We show 105 stars with $\delta P\!A \leqslant 10^\circ$, the same as in Figure \ref{fig:Sagittarius_PAgal}.
The two black arrows indicate the $P\!A$ of Planck's magnetic field inside the observed region (see Figure \ref{fig:Sagittarius_PAgal}, spatial resolution is set to $10\arcmin$).
The vertical dotted line represents the position angle for the Galactic Plane ($P\!A = 90^\circ$).
}
\label{fig:Sagittarius_PAgal_hist}
\end{figure}
The histogram of $P\!A$ shown in Figure \ref{fig:Sagittarius_PAgal_hist} shows a bimodal distribution centered around $30^\circ$ and $140^\circ$.
$P\!A = 90^\circ$, which is the direction parallel to the Galactic plane, corresponds to a minimum of the distribution.
Thus the observed $B_\mathrm{POS}$ is not parallel to but predominantly tilted from the Galactic plane.
The {\it P\!A}s and their distribution do not show particular variations or trends with sky coordinates (Figure \ref{fig:Sagittarius_PAgal}).

Figures \ref{fig:Sagittarius_PAgal} and \ref{fig:Sagittarius_PAgal_hist} also show Planck's observed magnetic field $P\!A$ for the same region (orange segments).
In the following, we refer to the polarimetry data observed by the Planck satellite at 353 GHz \citep[data release 3 (``PR3'');][]{2020A&A...641A...1P}, as provided by IRSA \citep{https://doi.org/10.26131/irsa558}, with a resolution set to $10\arcmin$.

Given that the Stokes parameters in the Planck data products are provided in the HEALPix convention rather than the IAU convention, we estimate the polarization position angle ({\it P\!A}) of the data using the following equation:
\[P\!A_{\mathrm{Planck}} = 0.5 \times \arctan2(-U_{\mathrm{Planck}}, Q_{\mathrm{Planck}}).\]
We estimate the {\it P\!A} of the magnetic field by rotating the polarization {\it P\!A} observed by Planck by $90^\circ$.
We will use the terms ``Planck's observed magnetic field'' or simply the ``Planck magnetic field'' for simplicity.
Similar to our stellar polarimetry data, the Planck magnetic field also shows $P\!A$ deviating from $90^\circ$ ($140^\circ\negthinspace.0$, $71^\circ\negthinspace.1$, $121^\circ\negthinspace.0$, and $66^\circ\negthinspace.0$ from north to south in Figure 1).
However, the angle offset from $90^\circ$ is generally larger for the stellar polarimetry magnetic field orientations.

\begin{figure}[tbp]
\centering
\includegraphics[width=\linewidth]{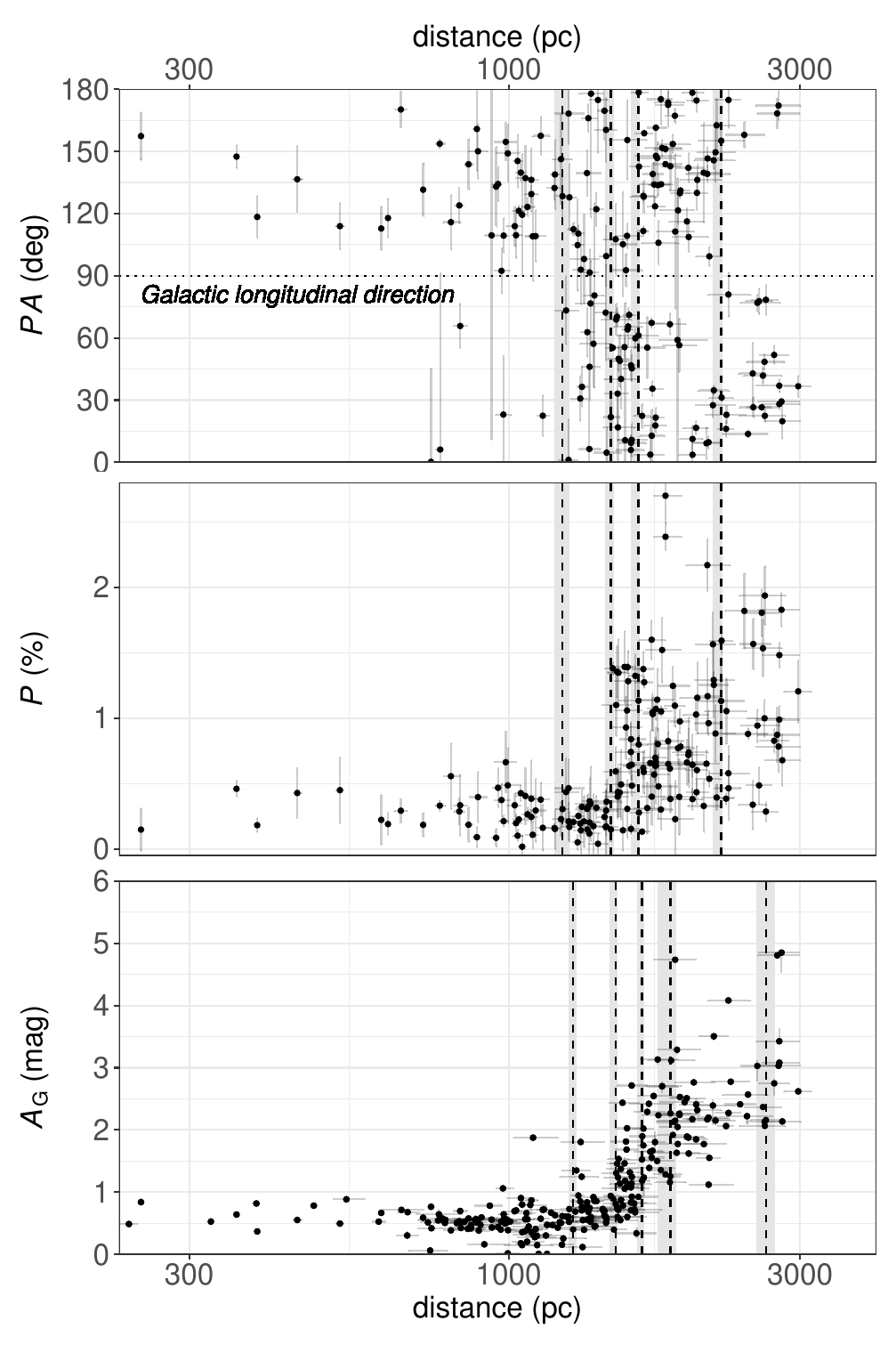}
\caption{
Distance dependence of polarimetry data ($P\!A$ and $P$; our observed 184 values), and $A_\mathrm{G}$ (the {\it Gaia} DR3 cataloged 259 values).
An observed $P\!A$ of $90^\circ$ indicates that the magnetic field is parallel to the Galactic plane.
The vertical dashed lines indicate the breakpoints of polarimetry data and $A_\mathrm{G}$ estimated by the breakpoint analysis.
Shaded areas correspond to 68\% confidence intervals (C.I.) of the estimation.
See text and \citet{2021ApJ...914..122D} for the details of the breakpoint analysis.
}
\label{fig:Sagittarius_qu_dist}
\end{figure}

The distance dependence of the optical polarimetry data is shown in Figure \ref{fig:Sagittarius_qu_dist}.
Specifically, we show how $P\!A$ and $P$ vary as a function of {\it Gaia} stellar distances estimated by \cite{2021AJ....161..147B}.
Note that $P\!A$ is mostly non-parallel to the Galactic plane, as shown in Figure \ref{fig:Sagittarius_PAgal_hist}.

Figure \ref{fig:Sagittarius_qu_dist} also shows interstellar extinction ($A_\mathrm{G}$) values taken from the {\it Gaia} DR3 catalog.
We find an apparent increase of about two mag in $A_\mathrm{G}$ at distances beyond $\sim 1.2$ kpc.
It further becomes $A_\mathrm{G} \geqslant 2.5$ mag beyond $\sim 2$ kpc.
We can attribute this increase in interstellar extinction at distances of about 1.2 -- 2 kpc to the dust in the Sagittarius arm.
The foreground component of $A_\mathrm{G} < 1$ mag (0.53 mag or $1.5 \times 10^{21}~\mathrm{H\mathchar`-atom~cm}^{-2}$ for $d < 1.23$ kpc, see Table \ref{tab:poleff}) can be attributed to the cloud(s) in the outskirt of the Aquila Rift at $d < 200$ pc (Section \ref{sec:intro}), and it is likely related to the Local Bubble shell \citep[][also see Figure \ref{fig:DR3cloud_BJ_2D}]{2019A&A...625A.135L,2020A&A...636A..17P}.

\subsection{Identification of Four Dust Clouds along the LOS using Breakpoint Analysis}
\label{sec:bpa}

\cite{2021ApJ...914..122D} showed that a breakpoint analysis, a statistical technique that detects the points at which data values make stepwise changes, can effectively recover the distance dependence of stellar polarimetry data.
Based on this breakpoint analysis, \cite{2021ApJ...914..122D} characterized the distribution of dust clouds as a function of distance along the LOS and the 3D structure of the magnetic field associated with those clouds.
The details of the breakpoint analysis are described in Appendix \ref{sec:breakpoint}.
We apply the breakpoint analysis to our observed $q_\mathrm{Gal}$ and $u_\mathrm{Gal}$, assuming a step change at each breakpoint and constant values between them, as was done by \cite{2021ApJ...914..122D}.
We identify four breakpoints, as shown in Table \ref{tab:breakpoints} (Polarimetry) and the dashed lines in Figure \ref{fig:Sagittarius_qu_dist} (top two panels), together with 68\% confidence intervals (C.I.) of the estimation.

We also perform the breakpoint analysis for $A_\mathrm{G}$ values similar to that for the polarimetry data.
The results are shown in Table \ref{tab:breakpoints} ($A_\mathrm{G}$) and Figure \ref{fig:Sagittarius_qu_dist} (the bottom panel).
We can find reasonable agreement between the two independent evaluations.
In particular, the three breakpoint distances on the near side show good consistency.
On the other hand, the $A_\mathrm{G}$ analysis finds an extra breakpoint at larger distances and these are roughly on either side of the polarimetry breakpoint.
We note that there are fewer stars with {\it Gaia}-estimated $A_\mathrm{G}$ values than those with polarimetry data in this distance range ($d > 1.6$ kpc; the number of stars in each distance range is listed in Table \ref{tab:constancy_check}).
Also, we find significant step change in $P\!A$ at 2.2 kpc (Figure \ref{fig:Sagittarius_qu_dist}).
As the focus of this paper is on the magnetic field structure inferred from the polarization data, we utilize the breakpoints detected in the polarization data analysis, which are expected to directly trace changes in the magnetic field structure, in the subsequent analyses.

\begin{table*}[t]
\begin{center}
\caption{Breakpoints estimated in $q_\mathrm{Gal},~u_\mathrm{Gal}$ and $A_\mathrm{G}$.}
\label{tab:breakpoints}
\begin{tabular}{llllll}
\hline
\hline
& \multicolumn{5}{c}{Breakpoints}\\
& \multicolumn{5}{c}{(pc)}\\
\hline
Polarimetry & $1225_{-34}^{+29}$ & $1470_{-28}^{+10}$ & $1632_{-42}^{+2}$ & \multicolumn{2}{c}{$2229_{-61}^{+3}$}\\
$A_\mathrm{G}$ & $1276_{-19}^{+10}$ & $1497_{-27}^{+3}$ & $1654_{-22}^{+8}$ & $1840_{-81}^{+33}$ & $2638_{-83}^{+83}$\\
\hline
\multicolumn{6}{p{0.50\textwidth}}{\bf Notes.}\\
\multicolumn{6}{p{0.50\textwidth}}{The error values indicate 68\% C.I. of the estimation. See text for the breakpoint analysis.}\\
\end{tabular}
\end{center}
\end{table*}

In the breakpoint analysis, as in \cite{2021ApJ...914..122D}, we assume that the values of $q_\mathrm{Gal}$ and $u_\mathrm{Gal}$ are both constant between the neighboring breakpoints.
To validate this assumption, we perform a linear fitting on each parameter between breakpoints to test if the slope is statistically consistent with a value of 0.
We perform the Student t-test for $q_\mathrm{Gal}$ and $u_\mathrm{Gal}$, and $A_\mathrm{G}$ as well.
Statistical $p$-values, which are for the null hypothesis that the slope of the distribution is equal to 0, are shown in Table \ref{tab:constancy_check}.
The null hypothesis that the slope of the distribution is equal to 0, i.e., a constant value, cannot be rejected as the $p$-values are all greater than 5\% for the tested cases.

\begin{table*}[t]
\begin{center}
\caption{Fitted parameters within each distance range.}
\label{tab:constancy_check}
\begin{tabular}{cccrrrrrr}
\hline
\hline
Distance range & \multicolumn{2}{c}{No. of stars} & \multicolumn{2}{c}{$q_\mathrm{Gal}$} & \multicolumn{2}{c}{$u_\mathrm{Gal}$} & \multicolumn{2}{c}{$A_\mathrm{G}$}\\
(kpc) & $q_\mathrm{Gal}$, $u_\mathrm{Gal}$ & $A_\mathrm{G}$ & \multicolumn{1}{c}{$\sigma^\mathrm{a}$} & \multicolumn{1}{c}{$p$-value$^\mathrm{b}$} & \multicolumn{1}{c}{$\sigma^\mathrm{a}$} & \multicolumn{1}{c}{$p$-value$^\mathrm{b}$} & \multicolumn{1}{c}{$\sigma^\mathrm{a}$} & \multicolumn{1}{c}{$p$-value$^\mathrm{b}$}\\
\hline
~~~~~~-- 1.23 & 45 & 104 & 0.46 & 0.647 & 0.55 & 0.585 & 0.03 & 0.977 \\
1.23 -- 1.47 & 29 & 42 & 0.37 & 0.708 & 0.14 & 0.889 & 0.67 & 0.503 \\
1.47 -- 1.63 & 25 & 39 & 0.25 & 0.803 & 0.18 & 0.859 & 0.69 & 0.487 \\
1.63 -- 2.23 & 61 & 54 & 0.88 & 0.378 & 0.05 & 0.957 & 1.43 & 0.153 \\
2.23 --~~~~~~ & 24 & 20 & 0.11 & 0.911 & 0.68 & 0.497 & 1.17 & 0.241 \\
\hline
\multicolumn{9}{p{0.52\textwidth}}{\bf Notes.}\\
\multicolumn{9}{p{0.6\textwidth}}{$^\mathrm{a}$Statistical deviation of the maximum likelihood slope value from 0.}\\
\multicolumn{9}{p{0.66\textwidth}}{$^\mathrm{b}$Statistical $p$-value for the null hypothesis that the slope of the distribution is equal to 0.
}\\
\end{tabular}
\end{center}
\end{table*}

Among the test results, the two distance ranges of $d \geqslant 1.63$ kpc, where the breakpoint estimate of $A_\mathrm{G}$ differs from that of the polarimetry data, the $A_\mathrm{G}$ values show smaller $p$-values.
However, the $p$-values are larger than 15\% and are still consistent with the assumption of constant $A_\mathrm{G}$ values at each distance range defined by the polarimetry.
Therefore, these analysis results can be considered as supporting evidence for the validity of the breakpoint analysis of the polarimetry data.

\begin{figure}[tp]
\centering
\includegraphics[width=\linewidth]{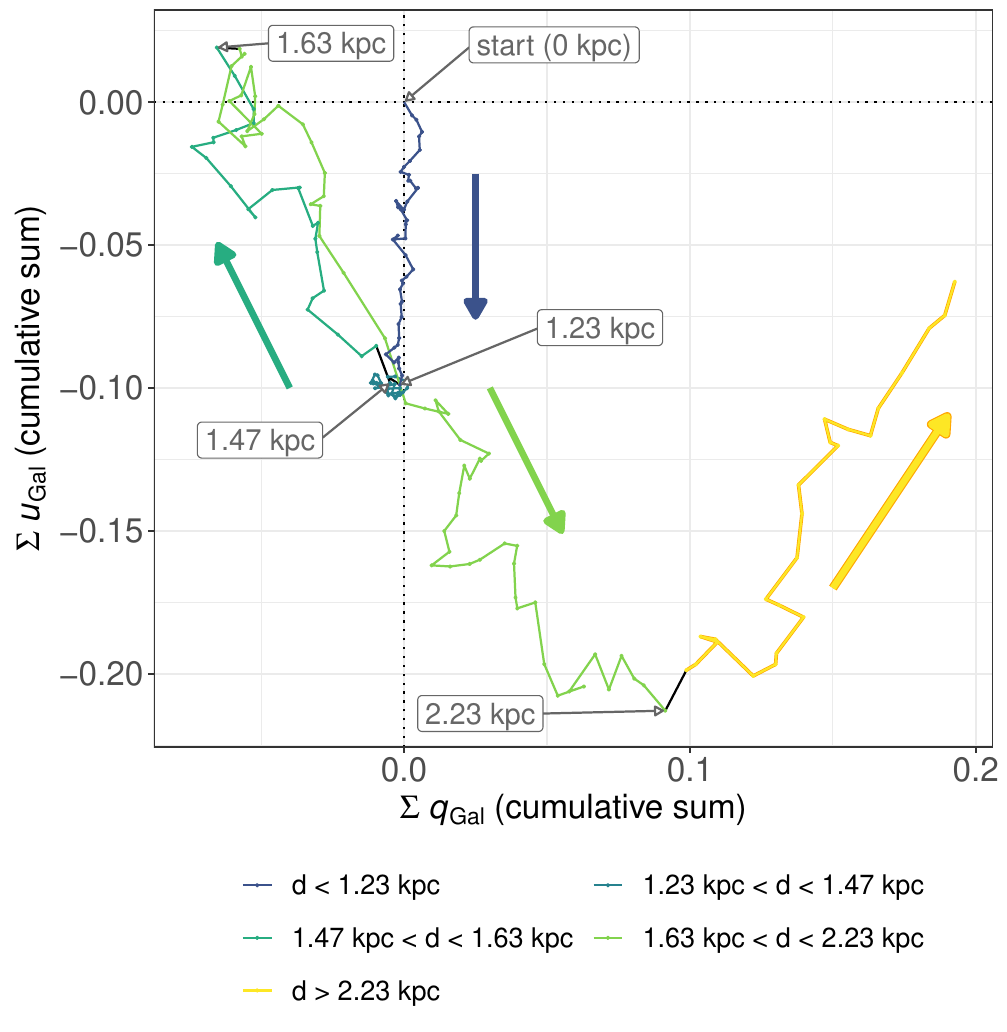}
\caption{
Cumulative sum plot of observed $q_\mathrm{Gal}$-$u_\mathrm{Gal}$ vectors, which are ordered by distance and shown as colored lines. The $q_\mathrm{Gal}$-$u_\mathrm{Gal}$ vectors are represented in fractions, with 0.1 corresponding to a polarization of 10\%. The labels indicate the stellar distances associated with the polarimetry breakpoints as listed in Table \ref{tab:breakpoints}. Additionally, the starting point of the sum vector is labeled as `start' at the origin. The cumulative sum vectors are clearly divided into five sections, consisting of four line segments and a clump between 1.23 kpc and 1.47 kpc. Colored arrows indicate the direction of the cumulative sum vector in each line segment.
}
\label{fig:Sagittarius_qu_cumsum}
\end{figure}

The constancy of the $q_\mathrm{Gal}$ and $u_\mathrm{Gal}$ values within each distance range implies that there is a discrete contribution of polarizing dust sheets at the breakpoints, while there is no significant contribution between breakpoints.
In Figure \ref{fig:Sagittarius_qu_cumsum}, we visually confirm this discrete polarization along the LOS by presenting a cumulative sum plot of the $q_\mathrm{Gal}$-$u_\mathrm{Gal}$ vectors with increasing distance.
In this plot, the sum vector defines a straight line while the position angles of the polarization remains constant if the position angles of the vectors are aligned.
This is because the vector sum averages out the random component of each vector.
On the other hand, if they are not aligned, a change in the polarization $P\!A$ turns the direction of the path of the cumulative sum plot.

As shown in Figure \ref{fig:Sagittarius_qu_cumsum}, the cumulative sum can be described by the combination of five sections, including four line segments and a clump between 1.23 kpc and 1.47 kpc.
The four line segments indicate that the $q_\mathrm{Gal}$-$u_\mathrm{Gal}$ vectors are well-aligned in each distance range.
The phase angle of the $q_\mathrm{Gal}$-$u_\mathrm{Gal}$ vector on the $\sum q_\mathrm{Gal} \mathchar`- \sum u_\mathrm{Gal}$ plane corresponds to twice the $P\!A$, and therefore, it should be noted that vectors pointing in opposite directions (e.g., the green and light green vectors in the figure) differ by $90^\circ$ in $P\!A$.
The clump between 1.23 kpc and 1.47 kpc shows that the length of $q_\mathrm{Gal}$-$u_\mathrm{Gal}$ vectors is 0 on average, which indicates that the vectors are aligned in one orientation in this distance range (due to the complete depolarization by its foreground cloud in this case).
In summary, Figure \ref{fig:Sagittarius_qu_cumsum} shows that the polarization vectors as a whole are well aligned in a specific direction for each of the five distance ranges, with discrete contributions of thin polarizing dust sheets at the breakpoints.
Different colors depict the distance range between the breakpoints, which correspond well to each line segment and clump.

The scattered distribution of dust clouds and their discrete contribution to the polarization (Figure \ref{fig:Sagittarius_qu_cumsum}) is comparable to the finding for the Perseus and (foreground) Taurus molecular clouds \citep{2021ApJ...914..122D} and the thin-layer model developed for high Galactic latitude clouds \citep{2023A&A...670A.164P}.
This suggests that the thin-layer model is also applicable to LOSs at low Galactic latitudes.
Therefore, in the following, we will assume that the discrete dust sheets/clouds at the four breakpoints, in addition to a foreground component before the first breakpoint, generate polarization in each distance range.
That is, `foreground', `1.23 kpc cloud', `1.47 kpc cloud', `1.63 kpc cloud', and `2.23 kpc cloud'.

Within the distance range where we identify dust clouds along the LOS ($d=1.2$ -- 2.2 kpc from the Sun), the vertical offset from the Galactic plane is $|Z| = 32$ -- 57 pc.
This vertical distance is comparable to or less than the scale height of the Galactic thin disk component \citep[50 -- 70 pc,][]{2006PASJ...58..847N,2007A&A...469..511K,2017ApJ...835...29Y} and well below that of the disk component of Galactic magnetic field models \citep[100 -- 400 pc,][]{2008A&A...477..573S,2012ApJ...757...14J,2013MNRAS.431..683J,2018ApJS..234...11H}.
Therefore, we are likely observing the Galactic disk component of the magnetic field.

\begin{figure}[tp]
\centering
\includegraphics[width=\linewidth]{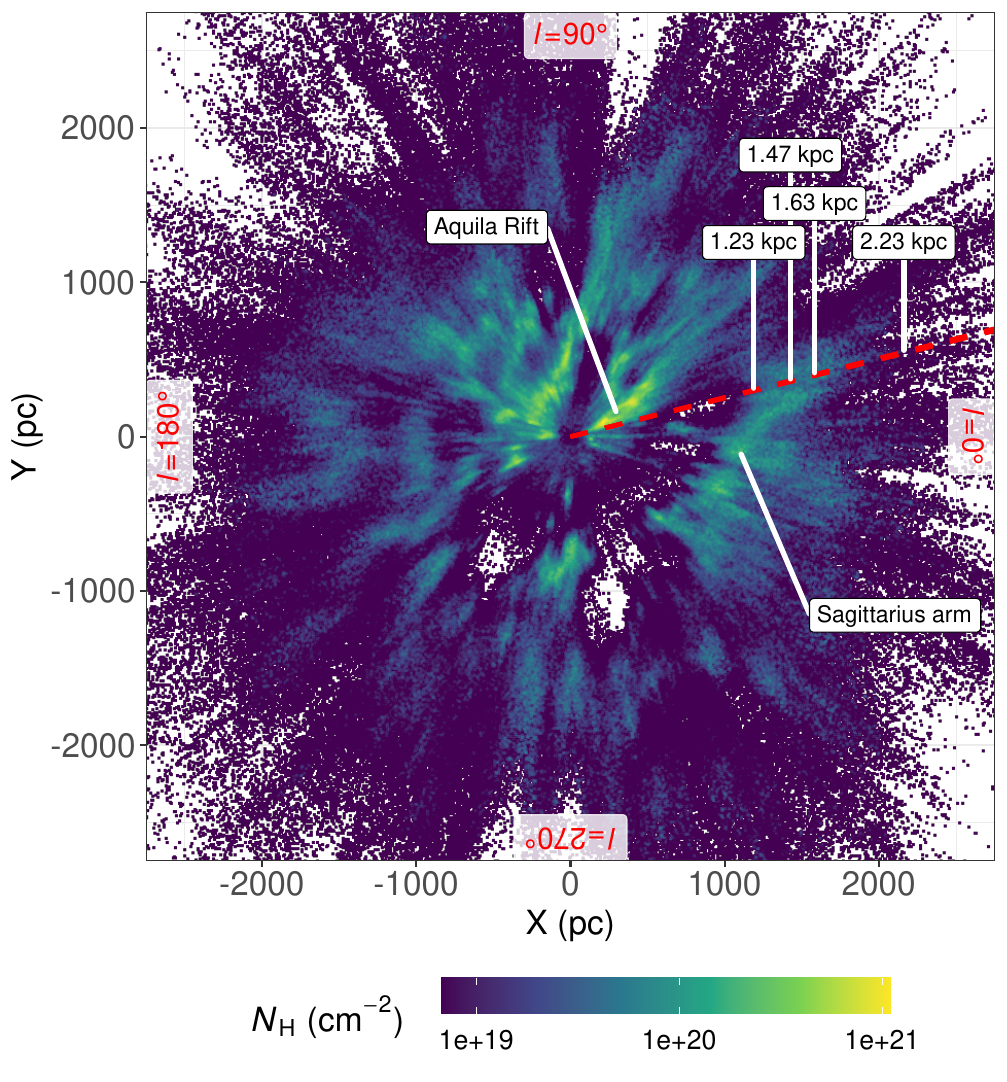}
\caption{
The red dashed line represents the sightline of the observation, while the positions of the four identified dust clouds (indicated by their distances) are shown on the Galactic plane.
The coordinates are heliocentric Galactic Cartesian coordinates, with the Sun located at the coordinate origin.
The X-axis points towards the Galactic center, the Y-axis points in the direction of Galactic rotation (Galactic plane at $l = 90^\circ$), and the Z-axis points towards the Galactic North pole (not depicted in the figure).
The color scale represents the surface density of the dust cloud within $Z=\pm100$ pc (see Appendix \ref{sec:surface_density} for the surface density estimation).
The regions of high dust surface density surrounding the 1.23 kpc, 1.47 kpc, and 1.63 kpc clouds correspond to the Sagittarius spiral arm.
}
\label{fig:DR3cloud_BJ_2D}
\end{figure}

We estimate the 3D distribution of dust clouds in the Galactic disc by applying the breakpoint analysis to $A_\mathrm{G}$ values, as described in Appendix \ref{sec:surface_density}. The color scale in Figure \ref{fig:DR3cloud_BJ_2D} represents the surface density of these dust clouds within a range of $\pm 100$ pc from the Galactic plane.
The red dashed line in the figure indicates the LOS of the observation. The positions of the four identified dust clouds along the LOS are indicated by their respective distances.

The high dust surface density structure observed around the 1.23 kpc, 1.47 kpc, and 1.63 kpc clouds in Figure \ref{fig:DR3cloud_BJ_2D} corresponds to the Sagittarius arm. The surface density around the 2.23 kpc cloud appears to be relatively low.
However, this does not necessarily imply the absence of dust clouds or the absence of the Sagittarius and Scutum arm structures, as dust clouds located on the far side within the Sagittarius arm may go undetected, being hidden behind the dust clouds on the near side of the Sagittarius arm or other foreground clouds.

In summary, our observations identify multiple clouds in the Sagittarius arm and detect polarization at each distance range.

\subsection{Magnetic Field Structure of Each Dust Cloud}
\label{sec:each_cloud}

\begin{figure}[tp]
\centering
\includegraphics[width=\linewidth]{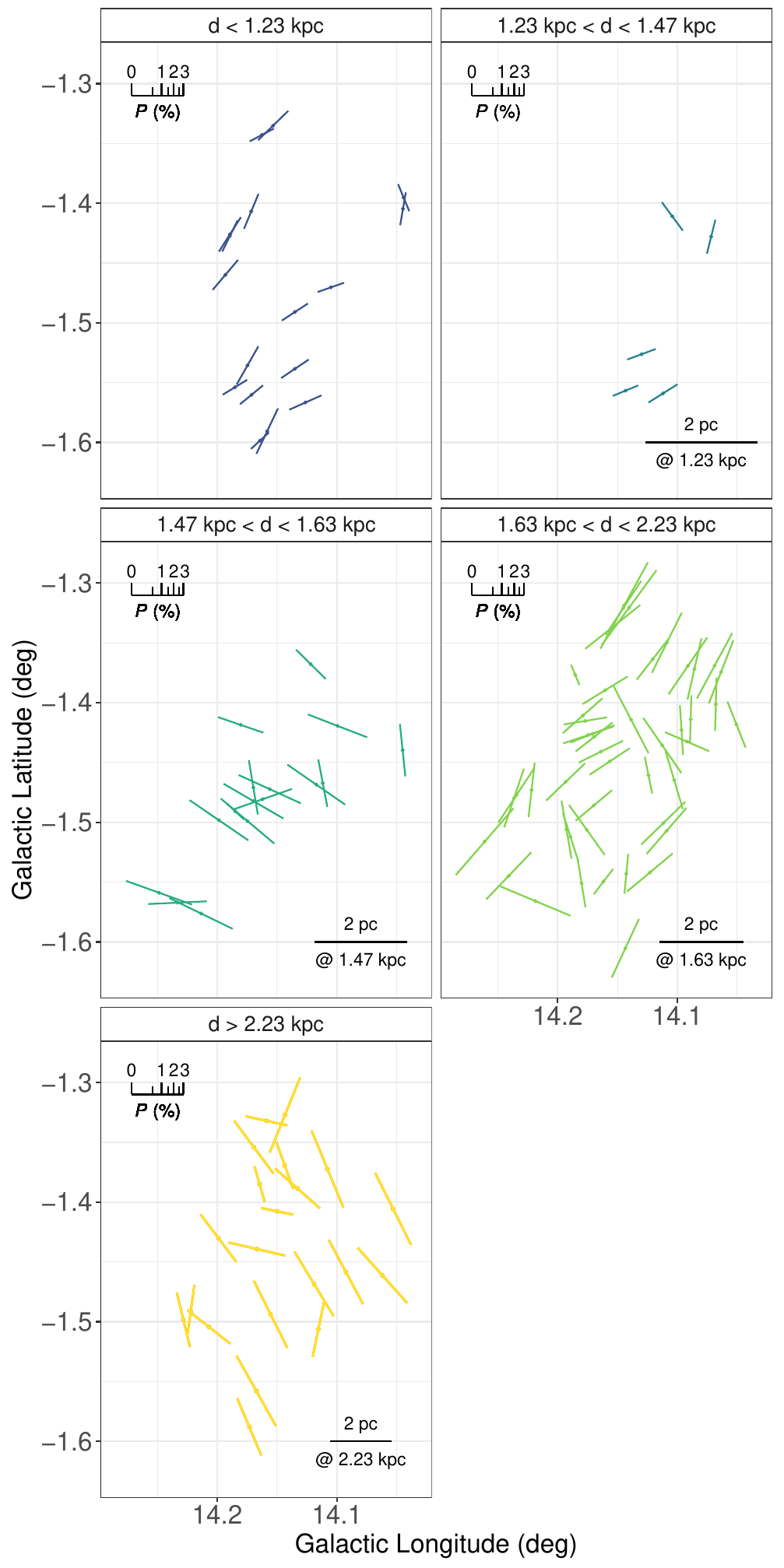}
\caption{
Spatial distribution of the observed position angles ($P\!A$) for each distance range.
Here we plot 105 data with uncertainties $\delta P\!A \leqslant 10^\circ$.
}
\label{fig:Sagittarius_qu_cloud}
\end{figure}

\begin{figure}[tp]
\centering
\includegraphics[width=\linewidth]{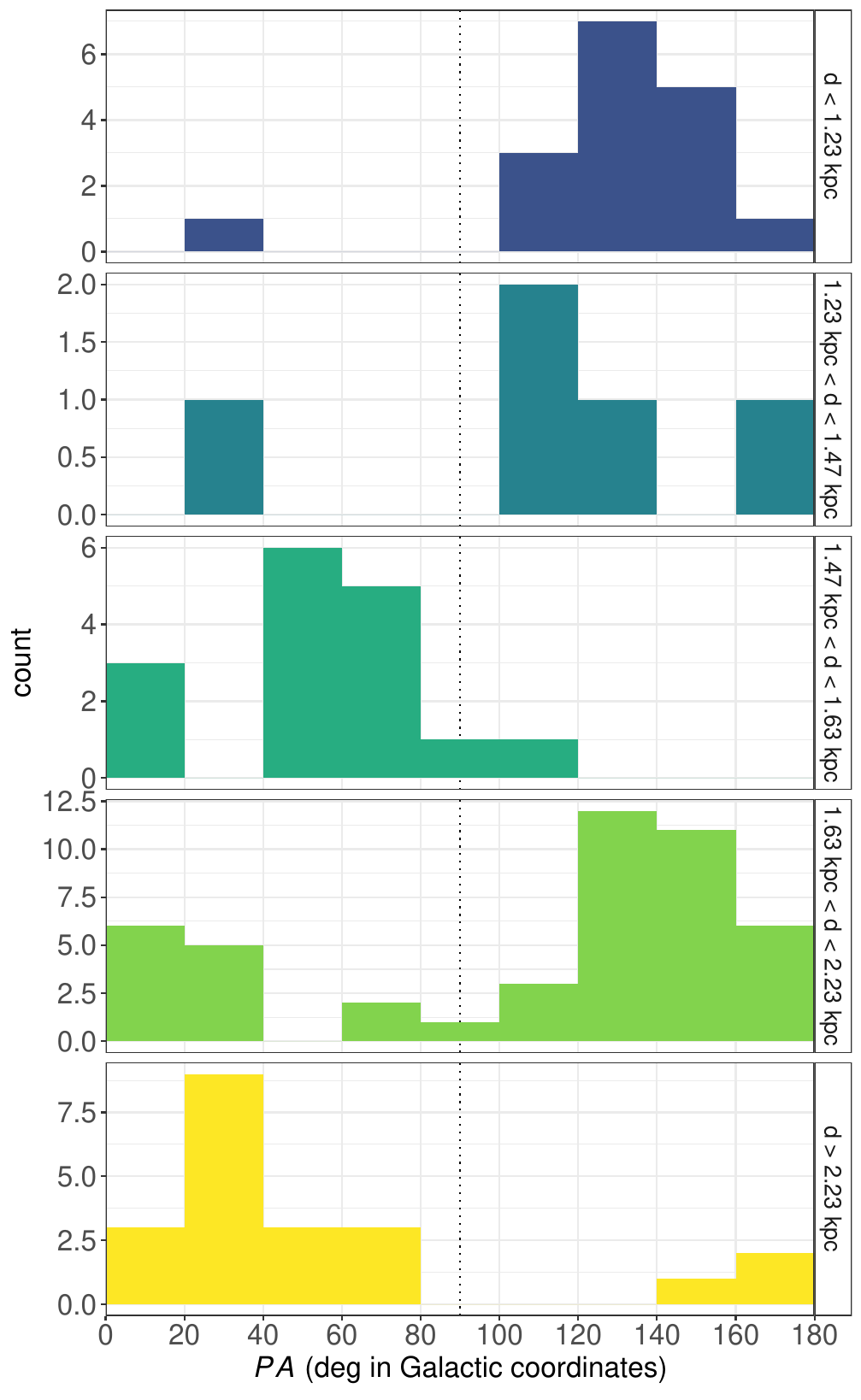}
\caption{
Histogram of the polarization angles ($P\!A$) for each distance range.
We plot data with uncertainties $\delta P\!A \leqslant 10^\circ$.
The bin width of the histograms is $20^\circ$.
The vertical dotted line indicates $P\!A = 90^\circ$ that is the position angle parallel to the Galactic plane.
}
\label{fig:Sagittarius_qu_cloud_hist}
\end{figure}

\begin{table*}[t]
\begin{center}
\caption{Angular mean and standard deviation of $P\!A$ within each distance range.}
\label{tab:cmean}
\begin{tabular}{ccrrrrrr}
\hline
\hline
Distance range & Cloud & \multicolumn{4}{c}{Circular mean} & \multicolumn{2}{c}{Circular SD $(\breve{\sigma}_{P\!A})$}\\
& & \multicolumn{2}{c}{Equatorial coordinates} & \multicolumn{2}{c}{Galactic coordinates} & &\\
& & \multicolumn{1}{c}{Observed} & \multicolumn{1}{c}{Intrinsic} & \multicolumn{1}{c}{Observed} & \multicolumn{1}{c}{Intrinsic} & \multicolumn{1}{c}{Observed} & \multicolumn{1}{c}{Differential}\\
(kpc) & & \multicolumn{1}{c}{(deg)} & \multicolumn{1}{c}{(deg)} & \multicolumn{1}{c}{(deg)} & \multicolumn{1}{c}{(deg)} & \multicolumn{1}{c}{(deg)} & \multicolumn{1}{c}{(deg)}\\
\hline
~~~~~~-- 1.23 & foreground & $72.6_{-2.9}^{+2.9}$ & $^\mathrm{a}\,72.6_{-2.9}^{+2.9}$ & $134.5_{-2.8}^{+2.8}$ & $^\mathrm{a}\,134.5_{-2.8}^{+2.8}$ & $29.3_{-2.3}^{+2.4}$ & $^\mathrm{a}\,29.3_{-2.3}^{+2.4}$ \\
1.23 -- 1.47 & 1.23 kpc & $22.9_{-49.0}^{+47.1}$ & $164.2_{-5.1}^{+5.1}$ & $85.2_{-48.3}^{+48.4}$ & $46.1_{-4.8}^{+4.7}$ & $63.2_{-6.9}^{+8.9}$ & $33.6_{-3.7}^{+4.1}$ \\
1.47 -- 1.63 & 1.47 kpc & $176.7_{-2.4}^{+2.5}$ & $176.2_{-3.1}^{+3.0}$ & $58.6_{-2.2}^{+2.2}$ & $58.1_{-2.8}^{+2.8}$ & $34.5_{-2.6}^{+2.8}$ & $34.8_{-3.0}^{+3.1}$ \\
1.63 -- 2.23 & 1.63 kpc & $90.1_{-1.6}^{+1.6}$ & $88.3_{-1.5}^{+1.5}$ & $152.0_{-1.4}^{+1.5}$ & $150.2_{-1.4}^{+1.4}$ & $37.2_{-1.3}^{+1.4}$ & $22.3_{-1.4}^{+1.4}$ \\
2.23 --~~~~~~ & 2.23 kpc & $146.0_{-1.3}^{+1.4}$ & $158.4_{-1.3}^{+1.3}$ & $28.0_{-1.3}^{+1.3}$ & $40.3_{-1.2}^{+1.2}$ & $29.2_{-1.6}^{+1.6}$ & $22.4_{-1.5}^{+1.7}$ \\
\hline
\multicolumn{2}{c}{All} & $109.1_{-3.6}^{+3.6}$ & $^\mathrm{a}\,109.1_{-3.6}^{+3.6}$ & $171.0_{-3.3}^{+3.2}$ & $^\mathrm{a}\,171.0_{-3.3}^{+3.2}$ & $54.9_{-1.9}^{+2.1}$ & $^\mathrm{a}\,54.9_{-1.9}^{+2.1}$\\
\hline
\multicolumn{8}{p{0.81\textwidth}}{\bf Notes.}\\
\multicolumn{8}{p{0.81\textwidth}}{The raw observed values for each distance range are shown in the `Observed' column.
The mean intrinsic $P\!A$ for each cloud, calculated by subtracting the foreground contributions of all components in front of the respective cloud, is presented in the `Intrinsic' column.
The circular standard deviation (SD) values subtracting the foreground contribution are shown in the `Differential' column.
Note that they are biased by the variation caused by the observational error and not showing the intrinsic magnetic field angular variation of each cloud.
See text and \cite{2020ApJ...899...28D} for the definitions of the circular mean and the circular standard deviation.}\\
\multicolumn{8}{p{0.81\textwidth}}{$^\mathrm{a}$The `Intrinsic' and `Differential' values of the foreground cloud and the `All' values are the same as those of the observed values because there is no foreground component to be subtracted.}\\
\end{tabular}
\end{center}
\end{table*}

We show the distribution of polarization pseudo-vectors and their $P\!A$ for each distance range in Figures \ref{fig:Sagittarius_qu_cloud} and \ref{fig:Sagittarius_qu_cloud_hist}.
As in Figures \ref{fig:Sagittarius_PAgal} and \ref{fig:Sagittarius_PAgal_hist}, we plot only the data points with good $P\!A$ determination ($\delta P\!A \leqslant 10^\circ$), which corresponds to 105 objects.
The overall distribution of $P\!A$ in Figures \ref{fig:Sagittarius_PAgal} and \ref{fig:Sagittarius_PAgal_hist} appeared spatially uncorrelated with a large scatter.
However, if we plot the data independently for each distance bin, as shown in Figures \ref{fig:Sagittarius_qu_cloud} and \ref{fig:Sagittarius_qu_cloud_hist}, the polarization pseudo-vectors instead show a well-ordered pattern.

We present the mean orientation and angular dispersion of the $P\!A$ for each distance range in Table \ref{tab:cmean}, calculated using the circular mean and circular standard deviation. The circular mean and circular standard deviation (hereafter $\breve{\sigma}_{P\!A}$) account for the $180^\circ$ ambiguity of the polarization pseudo-vectors.
This approach allows for an unbiased estimation of the standard deviation of $P\!A$, even if the deviation exceeds $50^\circ$, and is capable of capturing a wider, though not infinite, range of deviations in the $P\!A$ measurements compared to the usual arithmetic standard deviation, which saturates at $\pi / \sqrt{12}\,\mathrm{(rad)} = 51^\circ\negthinspace.96$ \citep{2020ApJ...899...28D}\footnote{Note that their definitions take into account $180^\circ$ ambiguity of the pseudo-vectors, unlike the nominal circular mean and the circular standard deviation that take into account $360^\circ$ ambiguity of the nominal vectors. See the definitions in their Appendix C.}.

We utilize all 184 objects selected according to the criteria described in Section \ref{sec:GaiaID}, including those with large $\delta P\!A$, for estimating the circular mean and $\breve{\sigma}_{P\!A}$.
This is in contrast to Figures \ref{fig:Sagittarius_PAgal}, \ref{fig:Sagittarius_PAgal_hist}, \ref{fig:Sagittarius_qu_cloud} and \ref{fig:Sagittarius_qu_cloud_hist}, which display data from only 105 objects.
To estimate the uncertainty of each parameter, we perform 10,000 Monte Carlo simulations. In each simulation, we add Gaussian random errors independently to the relative Stokes parameters {\it q} and {\it u} based on their respective uncertainties. From the generated samples, we calculate {\it P} and {\it PA} and obtain the required quantities for the analysis.
We show the median value of the 10,000 estimates as the maximum likelihood value and the 15.9\% and 84.1\% quantiles as the negative and positive errors in Table \ref{tab:cmean} and the following estimations in this paper.

The angular dispersion ($\breve{\sigma}_{P\!A}$) of the observed polarization pseudo-vectors are found in the `Observed' column in Table \ref{tab:cmean}.
Except for the 1.23 -- 1.47 kpc distance range, where the polarization pseudo-vectors are almost zero length due to the geometrical depolarization, the angular dispersion for each distance range is significantly smaller than that of the total data, confirming that the polarization pseudo-vectors of each distance bin are better aligned.

To accurately evaluate the magnetic field structure associated with each cloud, it is important to consider that the observed polarization is a result of integrating all contributions along the optical path to the stars. The relative Stokes parameters $q_\mathrm{Gal}$ and $u_\mathrm{Gal}$ can be approximated as an addition of the contributions from each element along the LOS, particularly in the case of low polarization levels \citep[say, $\ll 10$\%; e.g.,][]{2010A&A...510A.108P,2019ApJ...872...56P,2023A&A...670A.164P}. By subtracting the foreground contribution from the observed polarization in each distance range, we can obtain a more reliable approximation of the intrinsic magnetic field structure associated with each cloud. This allows us to isolate the specific magnetic field characteristics within each cloud, independent of the foreground effects.

The observed $q_\mathrm{Gal}$ and $u_\mathrm{Gal}$ data for the $n^\mathrm{th}$ distance range on the LOS are the sum of the contributions from all distance ranges from the $1^\mathrm{st}$ to the $n^\mathrm{th}$ distance ranges.
Similarly, the observed $q_\mathrm{Gal}$ and $u_\mathrm{Gal}$ data for the $(n-1)^\mathrm{th}$ distance range are the sum of the contributions from the $1^\mathrm{st}$ to the $(n-1)^\mathrm{th}$ distance ranges.
Therefore, to obtain the $q_\mathrm{Gal}$ and $u_\mathrm{Gal}$ values of the $n^\mathrm{th}$ distance range, we can subtract the $(n-1)^\mathrm{th}$ data from the $n^\mathrm{th}$ data, i.e., we can differentiate the observed $q_\mathrm{Gal}$ and $u_\mathrm{Gal}$ values of each distance range.

\begin{figure}[tp]
\centering
\includegraphics[width=\linewidth]{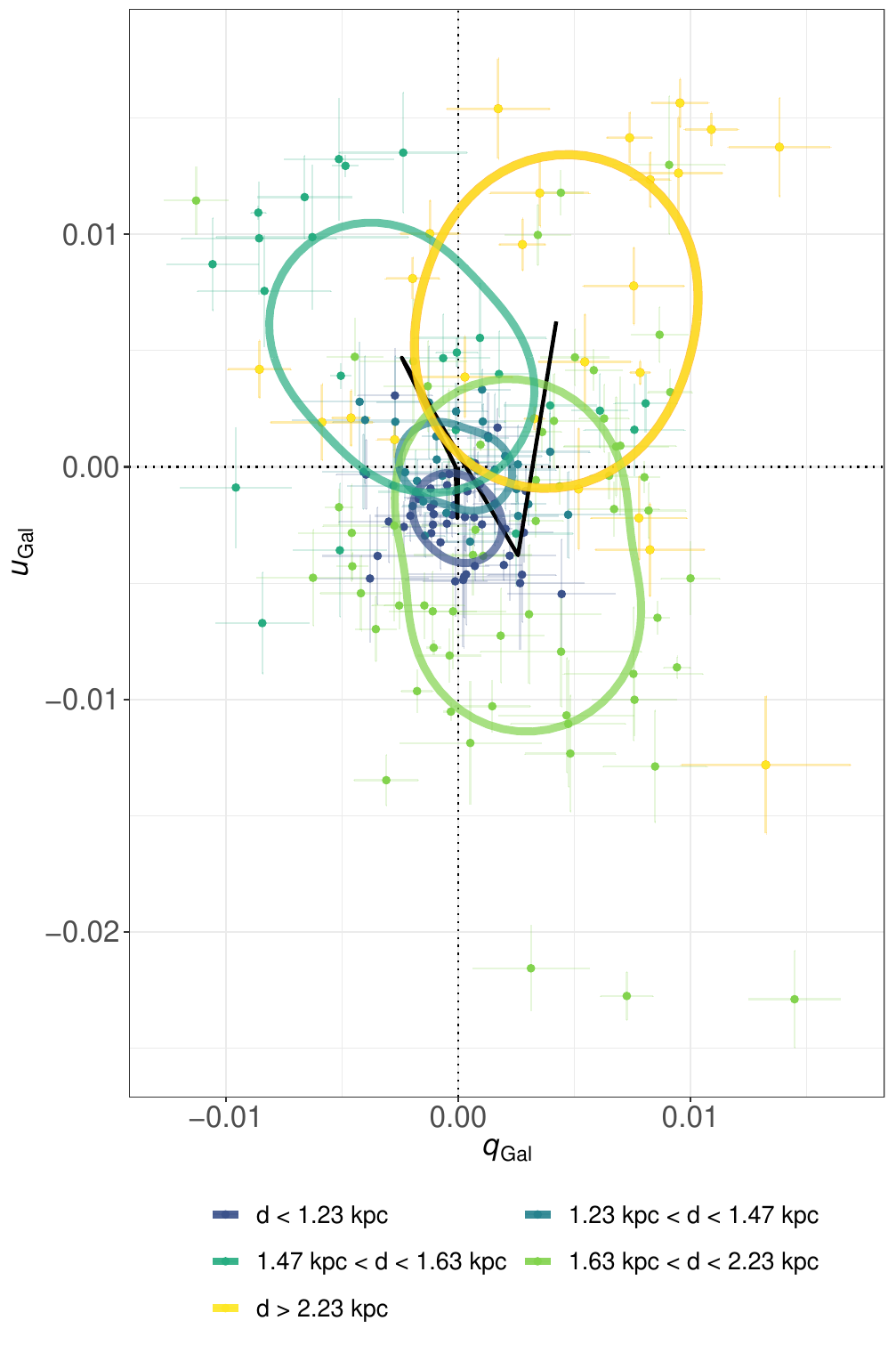}
\caption{
Distribution of $q_\mathrm{Gal}$-$u_\mathrm{Gal}$ by distance range.
Colored contours are the 1-$\sigma$ contours of the $q_\mathrm{Gal}$-$u_\mathrm{Gal}$ data scatter for each distance range.
The black line segments connect the average $q_\mathrm{Gal}$-$u_\mathrm{Gal}$  values of individual distance ranges and indicate the intrinsic polarization of each cloud.
}
\label{fig:Sagittarius_qu_trans}
\end{figure}

Figure \ref{fig:Sagittarius_qu_trans} shows the $q_\mathrm{Gal}$-$u_\mathrm{Gal}$ data distribution for all distance groups.
We also plot the 1-$\sigma$ contours of the $q_\mathrm{Gal}$-$u_\mathrm{Gal}$ data scatter for each distance range.
We can see that the data are discriminated by distance.
We estimate the average intrinsic polarization of each interstellar cloud by subtracting the average observed data of the immediately preceding cloud from the average observed data of a particular cloud.
The average intrinsic polarization vector is represented by each black line segment in Figure \ref{fig:Sagittarius_qu_trans}.
For each data point, similarly, we can obtain a better approximation of the $q_\mathrm{Gal}$ and $u_\mathrm{Gal}$ values of individual clouds by subtracting the average values of $q_\mathrm{Gal}$ and $u_\mathrm{Gal}$ of the immediately preceding cloud, which represents the integration of the contributions of foreground clouds.
The subtraction of the foreground contributions is thus equivalent to shifting the coordinate origin of the $q_\mathrm{Gal}$-$u_\mathrm{Gal}$ plane to the average of the $q_\mathrm{Gal}$ and $u_\mathrm{Gal}$ values of the immediately preceding cloud.
We will discuss the connection between this shift of origin and an anti-correlation between $P$ and $\breve{\sigma}_{P\!A}$ in Section \ref{sec:spanti}.

\begin{figure}[tp]
\centering
\includegraphics[width=\linewidth]{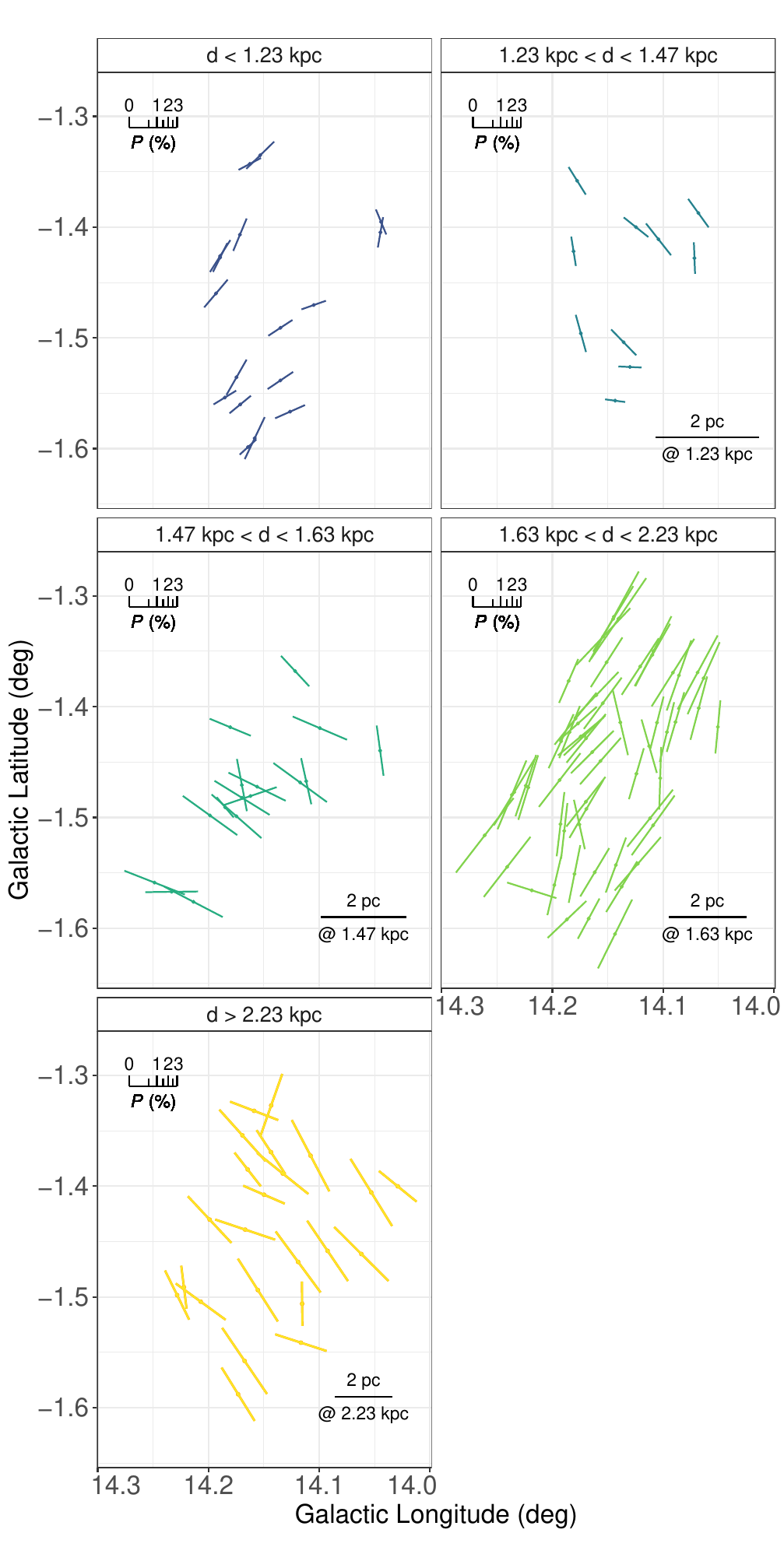}
\caption{
Same as shown in Figure \ref{fig:Sagittarius_qu_cloud}, but with the polarization pseudo-vectors of each cloud adjusted by subtracting the average foreground contribution and correcting for their average $P\!A$ and $P$ values. It is important to note that the individual pseudo-vectors in the figure are not corrected for the contribution of the foreground component to the variance of $P\!A$ and $P$; this correction can only be made statistically. Therefore, the vectors depicted in the figure are corrected only for their average values.
}
\label{fig:Sagittarius_qu_diff}
\end{figure}

\begin{figure}[tp]
\centering
\includegraphics[width=\linewidth]{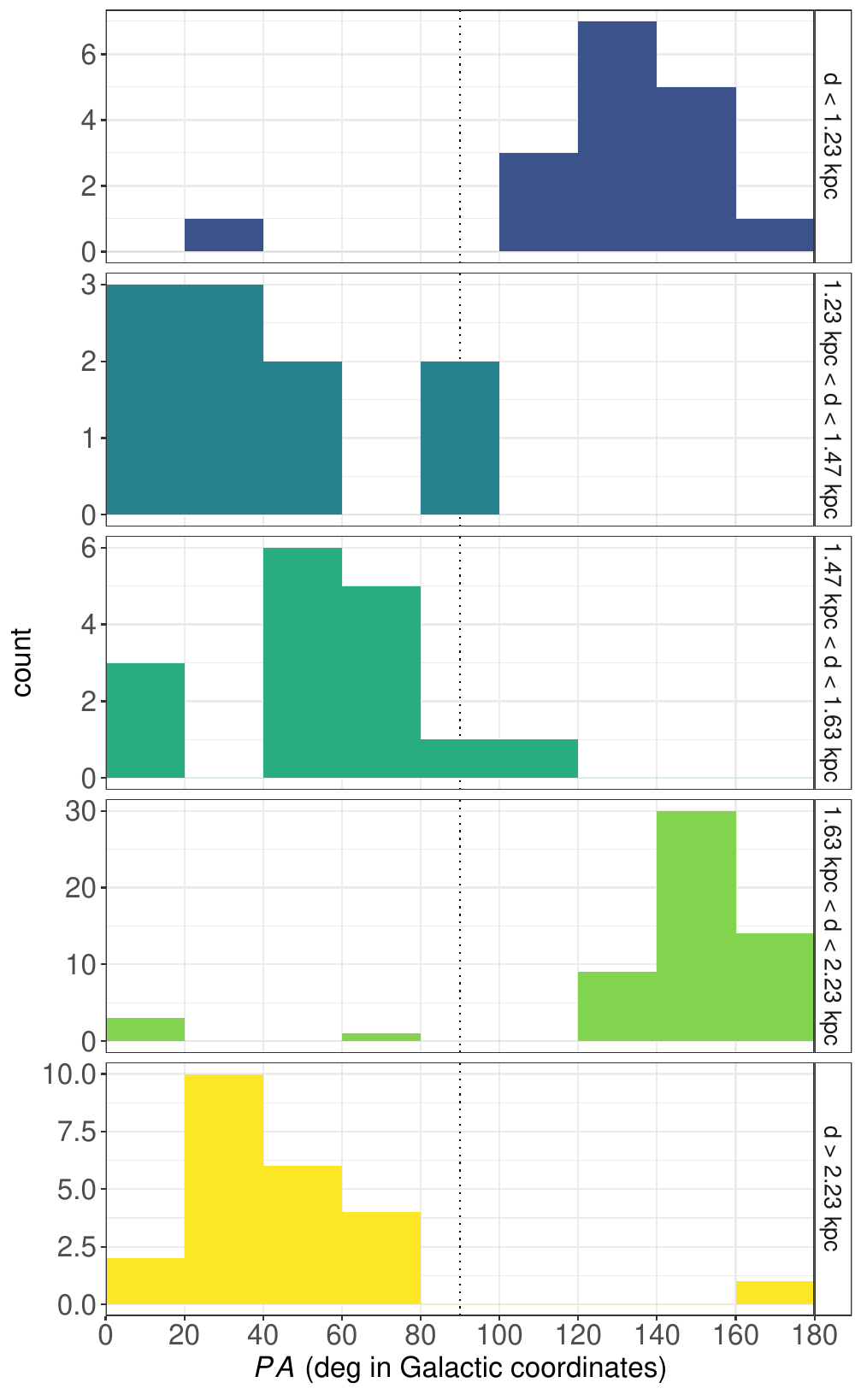}
\caption{
Same as shown in Figure \ref{fig:Sagittarius_qu_cloud_hist}, but with the $P\!A$ values of each cloud adjusted by subtracting the average foreground contribution.
}
\label{fig:Sagittarius_qu_diff_hist}
\end{figure}

Figures \ref{fig:Sagittarius_qu_diff} and \ref{fig:Sagittarius_qu_diff_hist} depict the distribution of polarization pseudo-vectors specific to each distance range, obtained by subtracting the mean foreground polarization. Comparing them to the raw observed values plotted in Figures \ref{fig:Sagittarius_qu_cloud} and \ref{fig:Sagittarius_qu_cloud_hist}, we observe that the polarization pseudo-vectors in each distance range exhibit better alignment.
This alignment enhancement can be attributed to the subtraction of the mean foreground polarization, which effectively shifts the origin of the $q_\mathrm{Gal}$-$u_\mathrm{Gal}$ plane to the average foreground value, as discussed earlier.
Consequently, this adjustment often elongates the $q_\mathrm{Gal}$-$u_\mathrm{Gal}$ vectors (resulting in increased $P$) and aligns them more coherently.
The improved alignment of these polarization pseudo-vectors indicates a well-ordered magnetic field associated with each dust cloud. 
The spatial scale of the observed region is approximately 5-10 pc, as indicated by the scales shown in Figure \ref{fig:Sagittarius_qu_diff}. This suggests that the spatial structure of the magnetic field associated with each cloud appears smooth at scales smaller than 5-10 pc, with a scale length of the magnetic field structure larger than 10 pc.

However, it is important to note that in Figure \ref{fig:Sagittarius_qu_diff}, we only subtract the mean foreground polarization, which means that the depicted vectors are corrected for the mean foreground contributions and not their variances. The contribution of the foreground component to the variance of $P\!A$ and $P$ can only be estimated statistically, and individual polarization pseudo-vectors cannot be corrected individually for this contribution.

Additionally, the observed variance of $P\!A$, or $(\breve{\sigma}_{P\!A})^2$, does not arise from a linear sum of contributions from each element along the line of sight (LOS), as will be discussed in Section \ref{sec:discussion}.
Moreover, the observed values of $\breve{\sigma}_{P\!A}$ are positively biased due to observation errors.
Instead, we compute the variance of $q_\mathrm{Gal}$-$u_\mathrm{Gal}$ vectors ($(\sigma_{q,u})^2$) specific to individual clouds by removing the foreground cloud's contribution as follows:
\begin{eqnarray}
\left(\sigma_{q,u,\text{intrinsic}}^n\right)^2 = \left[ \left(\sigma_{q,u,\text{observed}}^n\right)^2 - \left(\sigma_{q,u,\text{uncertainty}}^n\right)^2 \right] \nonumber \\ - \left[ \left(\sigma_{q,u,\text{observed}}^{n-1}\right)^2- \left(\sigma_{q,u,\text{uncertainty}}^{n-1}\right)^2 \right],~~~
\label{eq:sigma_intrinsic}
\end{eqnarray}
where the variance of $q_\mathrm{Gal}$-$u_\mathrm{Gal}$ for the $n$th cloud is denoted as $\left(\sigma_{q,u}^n\right)^2$, and the variance of $q_\mathrm{Gal}$-$u_\mathrm{Gal}$ for the immediately preceding cloud is denoted as $\left(\sigma_{q,u}^{n-1}\right)^2$.

Subsequently, this derived variance of $q_\mathrm{Gal}$-$u_\mathrm{Gal}$ vectors is employed to determine the variance in $P\!A$ specific to individual clouds.
For a more precise evaluation of the variance of $P\!A$, we will provide further discussion in Section \ref{sec:Bamp}.

Table \ref{tab:cmean} presents the circular mean and circular standard deviation of the polarization position angles for both the raw observed values (listed in the `Observed' columns) and the differential values. The differential values of the circular means are considered intrinsic to the magnetic field associated with each cloud, and we label these estimations as `Intrinsic' values in the table.

On the other hand, the differential values of $\breve{\sigma}_{P\!A}$ in Table \ref{tab:cmean} do not represent the angular dispersions specific to individual clouds, as explained previously.
Therefore, in Table \ref{tab:cmean}, we label the differential values of $\breve{\sigma}_{P\!A}$ as `Differential' instead of `Intrinsic'.

In the subsequent discussions, our primary focus will be on the intrinsic properties of the magnetic field associated with each cloud, unless stated otherwise.

\subsection{Polarization Fraction and Polarization Efficiency of Each Dust Cloud}
\label{sec:each_cloud2}

\begin{table*}[t]
\begin{center}
\caption{Polarization fraction and polarization efficiency within each distance range.}
\label{tab:poleff}
\begin{tabular}{ccrrc}
\hline
\hline
Cloud & Polarization fraction ($P$) & \multicolumn{1}{c}{$A_\mathrm{G}$} & \multicolumn{1}{c}{$N_\mathrm{H}^\mathrm{~a}$} & Polarization efficiency\\
& \multicolumn{1}{c}{(\%)} & \multicolumn{1}{c}{(mag)} & \multicolumn{1}{c}{($10^{21}\mathrm{cm}^{-2}$)} & \multicolumn{1}{c}{(\% mag$^{-1}$)}\\
\hline
foreground & $0.22_{-0.02}^{+0.02}$ & $0.53_{-0.00}^{+0.00}$ & $1.48_{-0.01}^{+0.01}$ & $0.42_{-0.04}^{+0.04}$ \\
1.23 kpc & $0.22_{-0.04}^{+0.04}$ & $0.17_{-0.01}^{+0.01}$ & $0.47_{-0.02}^{+0.02}$ & $1.35_{-0.23}^{+0.24}$ \\
1.47 kpc & $0.52_{-0.05}^{+0.05}$ & $0.41_{-0.01}^{+0.01}$ & $1.14_{-0.02}^{+0.02}$ & $1.28_{-0.12}^{+0.13}$ \\
1.63 kpc & $0.99_{-0.05}^{+0.05}$ & $0.98_{-0.01}^{+0.01}$ & $2.76_{-0.03}^{+0.03}$ & $1.00_{-0.05}^{+0.05}$ \\
2.23 kpc & $1.02_{-0.04}^{+0.04}$ & $0.75_{-0.02}^{+0.02}$ & $2.09_{-0.06}^{+0.05}$ & $1.37_{-0.06}^{+0.07}$ \\
\hline
All$^\mathrm{b}$ & $0.11_{-0.01}^{+0.01}$ & $1.15_{-0.00}^{+0.00}$ & $3.21_{-0.01}^{+0.01}$ & $0.10_{-0.01}^{+0.01}$ \\

\hline
\multicolumn{5}{p{0.25\textwidth}}{\bf Notes.}\\
\multicolumn{5}{p{0.65\textwidth}}{$^\mathrm{a}$ $N_\mathrm{H} = A_\mathrm{G} \cdot 2.21 \times 10^{21}/0.789$ is assumed \citep{2009MNRAS.400.2050G,2019ApJ...877..116W}.}\\
\multicolumn{5}{p{0.45\textwidth}}{$^\mathrm{b}$ Average of all the observed data.}\\
\end{tabular}
\end{center}
\end{table*}

Table \ref{tab:poleff} shows the polarization fraction ($P$) for each cloud.
To obtain these intrinsic $P$ values, we subtract the average observed $q_\mathrm{Gal}$ and $u_\mathrm{Gal}$ values of the immediately preceding cloud from the average observed $q_\mathrm{Gal}$ and $u_\mathrm{Gal}$ values of the specific cloud, and subsequently convert them into the polarization fraction ($P$).
This estimation can be visualized as the length of the black line segment in Figure \ref{fig:Sagittarius_qu_trans}. The average $P$ values of the raw observed data used for evaluating the intrinsic $P$ values are listed in Appendix \ref{sec:obs_values}.

To estimate the column density of each cloud, we utilize the {\it Gaia} DR3-cataloged interstellar extinction \citep[$A_\mathrm{G}$;][]{2022arXiv220606138A}. We calculate the average $A_\mathrm{G}$ within the ranges corresponding to each cloud and subtract the average $A_\mathrm{G}$ value of the immediately preceding cloud from the average $A_\mathrm{G}$ value of the specific cloud.

We estimate the column density ($N_\mathrm{H}$) of each cloud based on these $A_\mathrm{G}$ values, assuming $A_\mathrm{V} = A_\mathrm{G} / 0.789$ \citep[mag;][]{2019ApJ...877..116W} and $N_\mathrm{H} / A_\mathrm{V} = 2.21 \times 10^{21}$ \citep[H-atoms cm$^{-2}$ mag$^{-1}$;][]{2009MNRAS.400.2050G}. The estimated $A_\mathrm{G}$ and $N_\mathrm{H}$ values are presented in Table \ref{tab:poleff}. The average $A_\mathrm{G}$ and $N_\mathrm{H}$ values of the raw observed data used for evaluating these intrinsic $A_\mathrm{G}$ and $N_\mathrm{H}$ values can be found in Appendix \ref{sec:obs_values}.

The estimated intrinsic $A_\mathrm{G}$ of each cloud ranges from 0.17 -- 0.98 mag, corresponding to relatively low column densities of $N_\mathrm{H}\lesssim 2.76 \times 10^{21}~(\mathrm{H\mathchar`-atom~cm}^{-2})$.
This is because we have selected an observational field of view with relatively low interstellar extinction and with high accuracy measurements from {\it Gaia}'s optical trigonometry.
In other words, the observed magnetic field is not associated with star-forming regions within dense molecular clouds, but rather with the diffuse gas that likely surrounds the molecular gas in isolated clouds.
In fact, no corresponding CO molecular cloud is found in our field of view in catalogs \citep{2016ApJ...822...52R,2017ApJ...834...57M}, indicating that the gas is primarily atomic.
H{\small I} surveys \citep[e.g.,][]{2005A&A...440..775K,2015A&A...578A..78K} do not resolve the clouds due to low spatial and spectral resolutions, so the velocity dispersion of each dust cloud is unknown.
\citet{2020MNRAS.493..351C} identified dust clouds by referring to {\it Gaia} DR2 interstellar extinction data.
Their clouds No.\,505 at $l = 14^\circ\negthinspace.821,~b = -1^\circ\negthinspace.107$ and No.\,506 at $l = 13^\circ\negthinspace.370,~b = -0^\circ\negthinspace.212$ may correspond to our observed cloud(s) because of their spatial proximity to our FOV ($l = 14^\circ\negthinspace.15,~b = -1^\circ\negthinspace.47$).
The angular distances between the outer edges of their clouds and our FOV are $\sim 0.5^\circ$ (for the spatial extent of the clouds, see their Figures 505 and 506 available online\footnote{http://paperdata.china-vo.org/diskec/dustcloud/allcloud.pdf}).
The distance estimate of cloud No.\,505 is $1815.2 \pm 42.8$ pc and cloud No.\,506 is $1793.0 \pm 42.3$ pc.
The average distance of our detected four clouds (1.23 kpc, 1.47 kpc, 1.63 kpc, and 2.23 kpc), weighted by their column densities, is estimated to be $1767.0_{-1.3}^{+1.3}$ pc.
This average distance is almost identical to the distances of clouds No.\,505 and No.\,506.
We find more overlapping clouds in the LOS than in the literature, suggesting that we have detected tenuous dust clouds thanks to the distinct change in the magnetic fields' position angles as a function of distance.

We estimate the polarization efficiency \citep[e.g.,][]{2022dge..book.....W} by dividing $P$ by $A_\mathrm{G}$, as tabulated in Table \ref{tab:poleff}. The estimated polarization efficiency specific to individual clouds is 0.4$\%~\mathrm{mag}^{-1}$ for the foreground cloud and $1.0 \mathchar`-\mathchar`- 1.4\%~\mathrm{mag}^{-1}$ for the clouds in the Sagittarius arm. In a similar analysis, \cite{2021ApJ...914..122D} estimated a polarization efficiency of 1.5$\%~\mathrm{mag}^{-1}$ for the Taurus and Perseus molecular clouds. Taking into account the difference between the two observations \citep[0.7625 $\mu$m for Taurus and Perseus;][and 0.65 $\mu$m for this work]{1990ApJ...359..363G} and assuming a wavelength dependence of the fractional polarization as $P \propto \lambda^{-1.8}$ \citep{1990ARA&A..28...37M}, it corresponds to approximately 2.0$\%~\mathrm{mag}^{-1}$. Therefore, the observed efficiencies in our study are relatively smaller than those estimated for the Taurus and Perseus molecular clouds using the same method.

The pitch angle (the angle relative to the direction of the Galactic rotation) of the Sagittarius arm around the observed region is estimated to be $\psi\simeq 17^\circ$ \citep{2019ApJ...885..131R}.
Assuming the magnetic field follows the spiral arm structure, the magnetic field in the observation field is inclined to the POS by $i = 35^\circ$ (see Figure \ref{fig:DR3cloud_BJ_2D}).
In this case, the expected polarization fraction is approximately 0.7 times the maximum value, or $1.4\%\,\mathrm{mag}^{-1}$, if the maximum value is $\sim 2.0\%\,\mathrm{mag}^{-1}$ as observed in Taurus and Perseus clouds, based on the relation $P \propto \cos^2 i$.
That is, the smaller polarization efficiency found in the Sagittarius arm compared to Taurus and Perseus molecular clouds may be partly due to the tilted magnetic field orientation to the POS in the Sagittarius arm, while it is nearly parallel to the POS in Taurus and Perseus \citep[e.g.,][]{2012ApJ...757...14J}.
The difference in polarization efficiency of clouds in the Sagittarius arm may indicate that the magnetic field structure in the arm has a substantial variation in the in-plane direction of the Galaxy in addition to the direction perpendicular to the Galactic plane.
This variation in polarization efficiency may arise from a combination of factors, including differences in the alignment of dust particles and the intricate geometry of the magnetic field.

\section{Discussion} \label{sec:discussion}

\subsection{Anti-Correlation between Position Angle Dispersion and Polarization Fraction}
\label{sec:spanti}

\begin{figure}[tp]
\centering
\includegraphics[width=\linewidth]{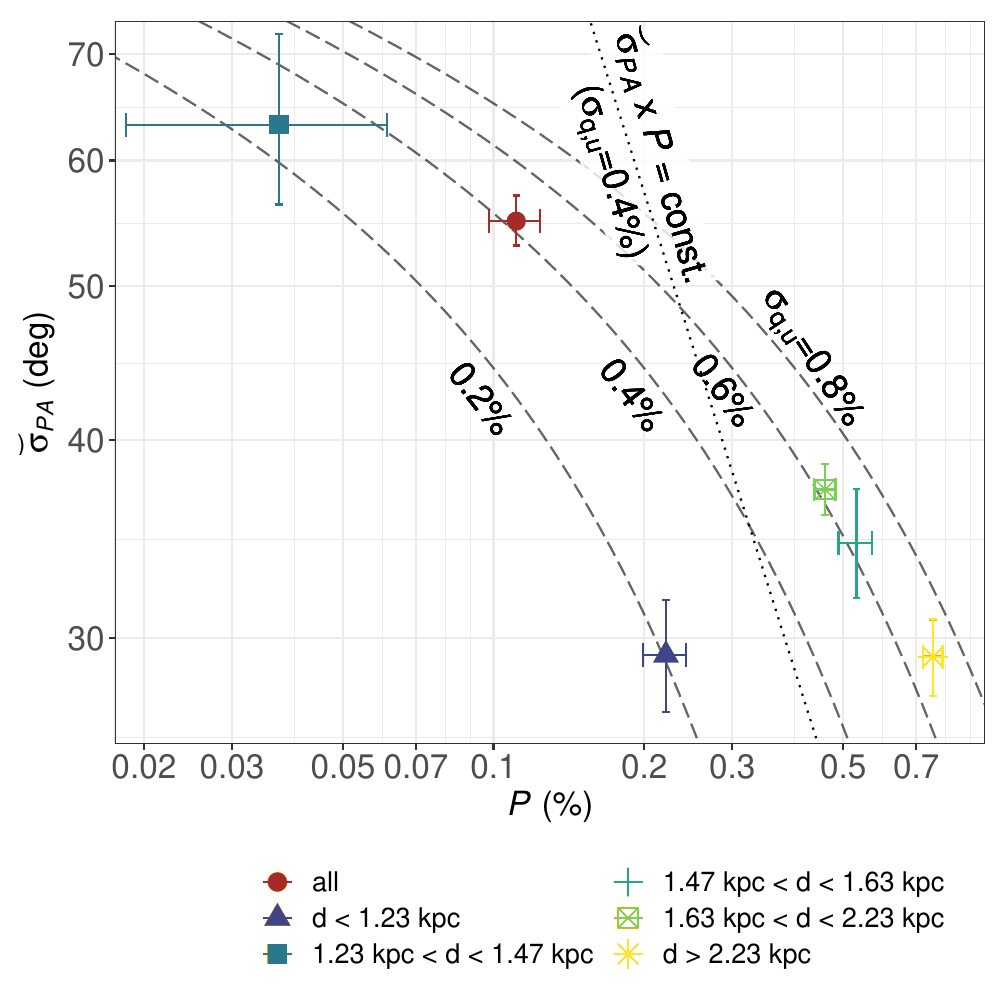}
\caption{
Relationship between the observed values of the polarization fraction ($P$) and the polarization angle dispersion ($\breve{\sigma}_{P\!A}$) in our samples.
The dashed lines represent the relationship between the position angle dispersion and the $P$ value, based on the marginal distribution of $P\!A$ described by Equation (\ref{eq:norm2d}).
These lines are plotted for $\sigma_{q,u}$ values of 0.8\%, 0.6\%, 0.4\%, and 0.2\%, representing different levels of dispersion.
The diagonal dotted line shows the approximation of the theoretical $P$-dependence, which corresponds to the $\breve{\sigma}_{P\!A} \propto P^{-1}$ correlation pointed out by \citet{2020A&A...641A..12P}, in the case of $\sigma_{q,u} = 0.4\%$ (see also Figure \ref{fig:norm2d}).
}
\label{fig:s-p}
\end{figure}

\citet{2020A&A...641A..12P} reported an anti-correlation between the dispersion of polarization angles and the polarization fraction \citep[also see][]{2016ApJ...824..134F}.
They attributed this anti-correlation to variations in the magnetic field structure along the LOS.
Figure \ref{fig:s-p} illustrates the mean observed polarization fraction ($P$) and position angle dispersion ($\breve{\sigma}_{P\!A}$) estimated for each distance range in the optical polarimetry data. These values, presented as the `Observed' values of $\breve{\sigma}_{P\!A}$ and $P$ in Tables \ref{tab:cmean} and \ref{tab:poleff}, represent measurements of multiple magnetic field components superimposed along the LOS at their respective distances. In other words, Figure \ref{fig:s-p} showcases the relationship between $P$ and $\breve{\sigma}_{P\!A}$ associated with different numbers of magnetic field layers along the LOS.

The data presented in Figure \ref{fig:s-p} show an anti-correlation, albeit with a slightly shallower slope compared to the correlation reported by \citet{2020A&A...641A..12P} as $\breve{\sigma}_{P\!A} \times P = \mathrm{const.}$
In the following analysis, we will investigate whether this shallower anti-correlation can be attributed to the same correlation reported in \citet{2020A&A...641A..12P}.

\subsubsection{Theoretical Curve}

A geometrical depolarization caused by multiple magnetic field layers along the LOS is equivalent to shifting the coordinate origin of the $q$-$u$ plane, as discussed in Section \ref{sec:each_cloud}.
When the origin of the coordinate system deviates further from the distribution of $q$ and $u$ data, the polarization fraction $P$ increases proportionally.
At the same time, the polarization position angle dispersion $\breve{\sigma}_{P\!A}$ decreases approximately inversely, particularly when $P$ is sufficiently large.
This dependence of $\breve{\sigma}_{P\!A}$ on $P$ is the same as that of the estimation error of $P\!A$ derived from the observed $q_\mathrm{Gal}$ and $u_\mathrm{Gal}$ when the standard deviations of $q_\mathrm{Gal}$ and $u_\mathrm{Gal}$ ($\sigma_q,~\sigma_u$) are interpreted as uncertainties in $q_\mathrm{Gal}$ and $u_\mathrm{Gal}$, respectively, rather than standard deviations.
For isotropic uncertainty distributions where $\sigma_q \approx \sigma_u \equiv \sigma_{q,u}$, the marginal probability distribution $G$ of $P\!A$ can be expressed as \citep{1993A&A...274..968N,2012A&A...538A..65Q}:
\begin{equation}
\begin{split}
G(P\!A~&|~P_0,P\!A_0,\sigma_{q,u})\\
= & \frac{1}{\sqrt{\pi}}\left( \frac{1}{\sqrt{\pi}} + \eta_0 e^{\eta_0^2} [1 + \mathrm{erf}(\eta_0)] \right) e^{- \frac{P_0^2}{2\sigma_{q,u}^{2}}}, \\
\mathrm{where}&~\eta_0 = \frac{P_0}{\sqrt{2}\sigma_{q,u}} \cos \left[ 2(P\!A - P\!A_0) \right].
\end{split}
\label{eq:norm2d}
\end{equation}
Here, $P_0$ and $P\!A_0$ represent the average values of $P$ and $P\!A$, respectively, and ``erf'' denotes the Gaussian error function.

We can estimate the angular dispersion $\breve{\sigma}_{P\!A}$ from the probability distribution of ${P\!A}$ based on the function $G$ (hereafter $\breve{\sigma}_{G(P\!A)}$), for each value of $\sigma_{q,u}$, or more precisely, for each value of $P_0/\sigma_{q,u}$ (see Equation \ref{eq:norm2d}).
Since we cannot solve the function $G$ analytically, we numerically estimate the dependence of $\breve{\sigma}_{G(P\!A)}$ on $P$, shown in Figure \ref{fig:s-p}.
The dashed lines in Figure \ref{fig:s-p} show the $\breve{\sigma}_{G(P\!A)}$ dependence on $P$ for several example $\sigma_{q,u}$ values.
We observe a general agreement between the angular dispersion $\breve{\sigma}_{P\!A}$ obtained from observations and the theoretical $\breve{\sigma}_{G(P\!A)}$ values within the range of $\sigma_{q,u}$ values ranging from 0.2\% to 0.8\%.

\begin{figure}[tp]
\centering
\includegraphics[width=\linewidth]{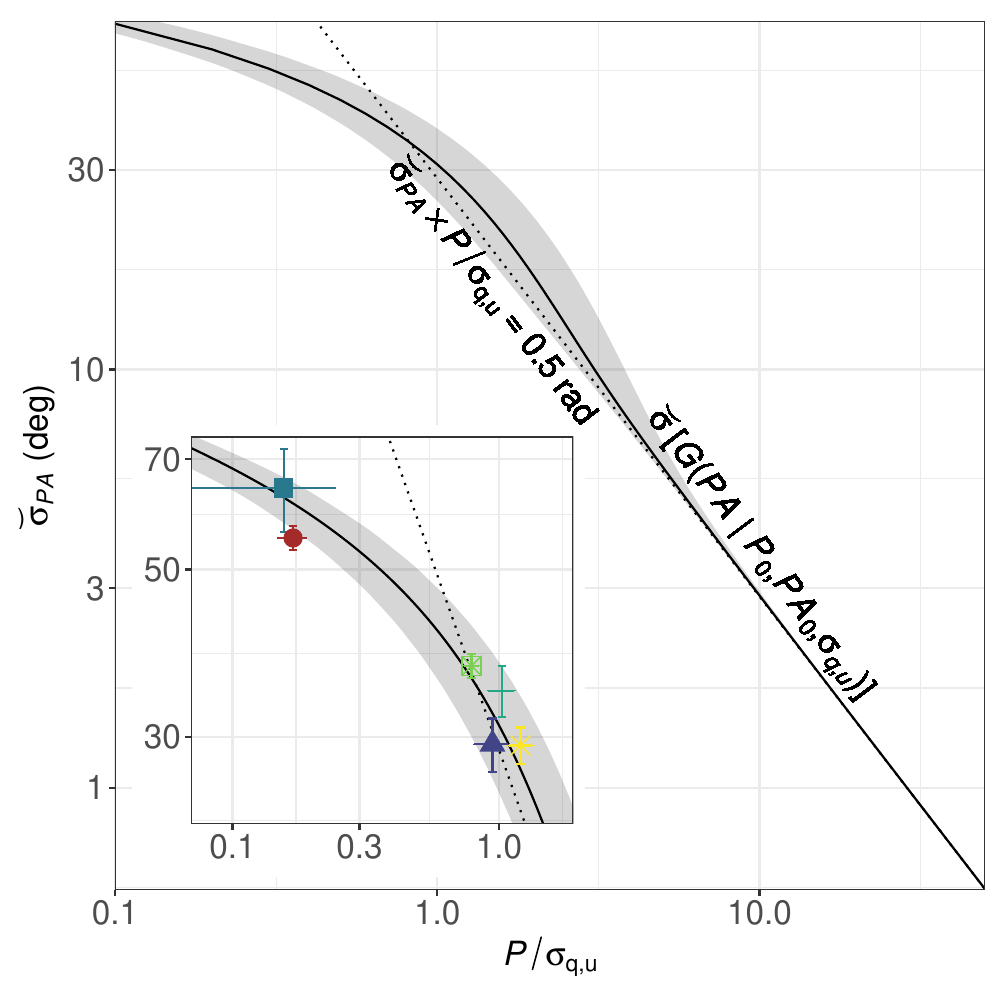}
\caption{
The solid line indicates the $P$ dependence of $\breve{\sigma}_{G(P\!A)}$, the same as in Figure \ref{fig:s-p} but with $P$ normalized by $\sigma_{q,u}$ ($P/\sigma_{q,u}$).
The dotted line shows the $\breve{\sigma}_{P\!A} \propto P^{-1}$ correlation pointed out by \citet{2020A&A...641A..12P}.
Inset is the distribution of the observed optical polarimetry data, whose symbols are the same as in Figure \ref{fig:s-p}.
The shaded area illustrates how the dependence of $\breve{\sigma}_{P\!A}$ on $P/\sigma_{q,u}$ deviates from the theoretical curve of $\breve{\sigma}_{G(P\!A)}$ when $\sigma_{q,u}$ is not perfectly isotropic and the aspect ratio of its distribution is 1.54. Please refer to the main text for details.
}
\label{fig:norm2d}
\end{figure}

In the $q\text{-}u$ plane, the angular dispersion $\breve{\sigma}_{G(P\!A)}$ corresponds to the spread of $q\text{-}u$ data, measured in radians from the origin of the $q\text{-}u$ plane. This angle can be approximated by the tangent of $\sigma_{q,u}$ with respect to $P$. This is why $\breve{\sigma}_{G(P\!A)}$ in Equation (\ref{eq:norm2d}) is a function of $P_0/\sigma_{q,u}$. We illustrate the comparison between $P_0/\sigma_{q,u}$ and $\breve{\sigma}_{G(P\!A)}$ in Figure \ref{fig:norm2d}. When normalizing the mean observed polarization fraction ($P$) estimated for each distance range in the optical polarimetry data shown in Figure \ref{fig:s-p} by the values of $\sigma_{q,u}$ for the same distance range, this normalization removes the dependence of all the observed $\breve{\sigma}_{P\!A}$ values and the data points should fall on the same theoretical curve of $\breve{\sigma}_{G(P\!A)}$ represented by the solid line in Figure \ref{fig:norm2d}.

The theoretical curve of $\breve{\sigma}_{G(P\!A)}$ follows the relation $\breve{\sigma}_{P\!A} \times P/\sigma_{q,u} = 0.5$ radian when $P$ is sufficiently large and $\breve{\sigma}_{P\!A} \ll 10^\circ$.
This is because the phase angle standard deviation of the $q\mathchar`-u$ vectors in radians is approximately equal to the ratio between $\sigma_{q,u}$ and $P$ if $P$ is sufficiently large compared to $\sigma_{q,u}$. Thus, $\breve{\sigma}_{P\!A}$ is approximately $0.5 \times \sigma_{q,u} / P$ radians.
On the other hand, when $P$ is small and $\breve{\sigma}_{P\!A} \gg 10^\circ$, the slope of the theoretical curve becomes larger than -1 and closer to 0.

We present a comparison of the optical polarimetry data with $\breve{\sigma}_{G(P\!A)}$ after normalizing $P$ by $\sigma_{q,u}$ in the inset of Figure \ref{fig:norm2d}.
The observations show a general agreement with the theoretical $\breve{\sigma}_{G(P\!A)}$ curve.

\subsubsection{Influence of Non-Isotropic $\sigma_q$ and $\sigma_u$ Distributions}
\label{sec:nonuniformity}

Equation (\ref{eq:norm2d}) or the solid line in Figure \ref{fig:norm2d} assumes isotropic uncertainty distributions ($\sigma_q \approx \sigma_u \equiv \sigma_{q,u}$). However, the observed distributions of $\sigma_{q,u}$ are not perfectly isotropic (Figure \ref{fig:Sagittarius_qu_trans}). When the distribution is non-isotropic, the dependence of $\breve{\sigma}_{P\!A}$ on $P/\sigma_{q,u}$ deviates from the theoretical curve of $\breve{\sigma}_{G(P\!A)}$. Due to the small sample size in our observations (minimum of 24 objects, Table \ref{tab:breakpoints}), it is not possible to distinguish whether this bias stems from a physical background or from a bias in the observation sampling itself. In the following, we demonstrate that the deviation from an isotropic Gaussian distribution observed in the data has a minimal impact on the estimation of $\breve{\sigma}_{P\!A}$ and does not affect the discussions in this paper.

We note, however, that it is important for future observations to increase the sample size and determine the precise shape of the $\sigma_{q,u}$ distributions, as these distributions contain information about the position angle distribution of turbulent magnetic fields and the spatial variation of dust properties; the $\sigma_{q,u}$ value measured in the direction perpendicular to each cloud's mean $q\mathchar`-u$ vector (hereafter $\sigma_{q,u\perp}$) approximates $\breve{\sigma}_{P\!A}$ and reflects the {\it P\!A} dispersion of the magnetic field on the POS, and the $\sigma_{q,u}$ value measured in the direction parallel to each cloud's mean $q\mathchar`-u$ vector (hereafter $\sigma_{q,u\parallel}$) is considered to arise from the angular dispersion in the LOS direction of the magnetic field, as well as the fluctuations in polarization efficiency for each region within the polarizing cloud (such as variations in column density and dust alignment efficiency, also see \citealp{2023A&A...670A.164P}).

If we approximate the distribution of $\sigma_{q}$ and $\sigma_{u}$ as an ellipse and estimate the aspect ratio of the major and minor axes, the aspect ratio of the observed $\sigma_{q,u}$ distribution ranges from a minimum of 1.22 (for the $d > 2.23$ kpc distance range) to a maximum of 1.54 (for the $1.63$--$2.23$ kpc distance range). In Figure \ref{fig:norm2d}, we illustrate how the dependence of $\breve{\sigma}_{P\!A}$ on $P/\sigma_{q,u}$ deviates from the theoretical curve of $\breve{\sigma}_{G(P\!A)}$ when $\sigma_{q,u}$ is not perfectly isotropic, represented by the shaded area. We indicate the deviation corresponding to the maximum aspect ratio of the observed data (1.54).

When the major axis of the distribution aligns with $\sigma_{q,u\parallel}$, $\breve{\sigma}_{P\!A}$ is maximized, corresponding to the upper boundary of the shaded area. This is because the proportion of data closer to the $q\mathchar`-u$ coordinate origin increases. Conversely, when the major axis aligns with $\sigma_{q,u\perp}$, $\breve{\sigma}_{P\!A}$ is minimized, corresponding to the lower boundary of the shaded area. This is because the variability of data with distance from the $q\mathchar`-u$ coordinate origin decreases, reducing the proportion of data closer to this origin.
In cases where the major axis of the distribution is oblique to both $\sigma_{q,u\parallel}$ and $\sigma_{q,u\perp}$, an intermediate dependence is observed.
When $\breve{\sigma}_{P\!A} \ll 10^\circ$, the deviation from the theoretical curve due to the non-isotropic $\sigma_{q,u}$ distribution can be ignored.

In the inset of Figure \ref{fig:norm2d}, we normalize the observed $P$ with $\sigma_{q,u\perp}$, because $\sigma_{q,u\perp}$ closely approximates $\breve{\sigma}_{P\!A}$. The estimated values of $\sigma_{q,u\perp}$ are provided in Table \ref{tab:sigma_tan_obs} in Appendix \ref{sec:obs_values}. As shown in the figure, the data align well with the theoretical curve of $\breve{\sigma}_{G(P\!A)}$ within the expected range of deviation, which arises from the anisotropy of the observed $\sigma_q$ and $\sigma_u$ values.

The asymmetric distribution around the mean positions of $\sigma_{q}$ and $\sigma_{u}$ data generally offsets the value of $\breve{\sigma}_{P\!A}$ from the predicted $\breve{\sigma}_{G(P\!A)}$ based on Equation (\ref{eq:norm2d}). In cases where the data follows a non-Gaussian distribution, a non-zero kurtosis does not have an effect, but if non-zero skewness is present, it influences the estimation of $\breve{\sigma}_{P\!A}$.

In Appendix \ref{seq:anisotropy}, we further check the deviation from the theoretical $\breve{\sigma}_{G(P\!A)}$ curve caused by the non-isotropic $\sigma_{q}$ and $\sigma_{u}$ distribution including the oblique and skewed $\sigma_{q,u}$. We present a comparison of estimated values of $\breve{\sigma}_{P\!A}$ with and without consideration of the anisotropic distribution of $\sigma_{q}$ and $\sigma_{u}$.
Even when considering the anisotropic distribution, it is emphasized that the difference from the case without consideration falls within the range of estimated uncertainties.

According to the discussion above, we conclude that the observed optical polarimetry data are consistent with the theoretical curve of $\breve{\sigma}_{G(P\!A)}$ taking into account the influence of the non-isotropic $\sigma_{q,u}$ distribution.
In the following, we proceed with the discussion using the intrinsic $\breve{\sigma}_{P\!A}$ for each cloud, considering the influence of the non-isotropic distribution of $\sigma_{q,u}$.

\subsubsection{Anti-Correlation Induced by Superposition of Multiple Magnetic Field Layers}

We find that for the optical polarimetry data, $\breve{\sigma}_{G(P\!A)}$ is significantly larger than $10^\circ$, and the slope of the theoretical curve in Figure \ref{fig:norm2d} is shallower than -1. On the other hand, in the study by \citet{2020A&A...641A..12P}, they referred to the data with $\breve{\sigma}_{P\!A} \lesssim 10^\circ$ when discussing the anti-correlation, where the theoretical curve exhibits a linear anti-correlation with a slope of -1, which is consistent with their findings.
In their study, \citet{2020A&A...641A..12P} demonstrated a general agreement of their observed values with a single anti-correlation: $\sigma_{P\!A} \times P = 31~(\mathrm{deg \cdot \%)}$. This suggests that the value of $\sigma_{q,u}$ from Planck does not vary significantly across different observed sources and is estimated to be $\sigma_{q,u} = 1.08\%$. However, it is worth noting that there is an order of magnitude variation in the observed $\sigma_{P\!A} \times P$ values from Planck, which is comparable to the variation we observe in optical polarimetry, where $\sigma_{q,u}$ ranges from 0.2\% to 0.8\% (Figure \ref{fig:s-p}, also see Table \ref{tab:poleff_obs}).

According to the above discussion, we can interpret the anti-correlation of optical polarimetry data shown in Figure \ref{fig:s-p} and the anti-correlation of Planck data discussed by \citet{2020A&A...641A..12P} as a distribution that follows the same function, $\breve{\sigma}_{G(P\!A)}$.
In other words, the anti-correlation observed by Planck can be created by the variation of cloud superposition along the LOS that causes the variation of geometrical depolarization due to the superposition of multiple magnetic field components along the LOS.
Our observations thus suggest that this multi-component geometrical depolarization is likely the primary cause of the anti-correlation observed along the LOS in the Sagittarius arm,
which confirms the discussion by \citet{2020A&A...641A..12P}.

In the \citet{2020A&A...641A..12P} model, the intensity ratio between the turbulent magnetic field ($B_\text{turb}$, or different components of the magnetic field between layers) and the uniform component ($B_\text{unif}$) is 0.9, and the fluctuation of the turbulent magnetic field within the Planck beam is negligible.
As a result, the ${P\!A}$ differs significantly between layers in the LOS in their model, but a well-aligned magnetic field is required within a single layer.
We note that in our observation, the magnetic field of individual clouds (Figure \ref{fig:Sagittarius_qu_diff}) is well aligned with position angles that vary significantly from one cloud to another and are notably different from those observed by Planck. This alignment remains consistent even at scales less than $10\arcmin$, which approximately corresponds to the native resolution of Planck's polarization data.
This observation is also in line with the discussion by \citet{2020A&A...641A..12P}.

The smooth magnetic field structure of each cloud, even at spatial scales below those resolved by Planck observations, along with the significant variation in position angles from one cloud to another, suggests that for diffuse clouds with $N_\mathrm{H} \lesssim 3\times 10^{21}\,\mathrm{cm}^{-2}$ in this study, the discrepancies between Planck and stellar polarization are likely attributed to differences in probed distances rather than differences in beam sizes.
Planck captures the superposition of all ISM along the LOS, whereas stellar polarization only probes the ISM located in front of each individual star.

\subsection{Amplitude of Turbulent Magnetic Field}
\label{sec:Bamp}

\subsubsection{Intrinsic $\breve{\sigma}_{P\!A}$ Values of Each Cloud}

As described in Section \ref{sec:spanti}, $\sigma_{q,u}$ is a function of $\breve{\sigma}_{P\!A}$ and $P$ derived from Equation (\ref{eq:norm2d}) and can be approximated as $\sigma_{q,u} \simeq ( \breve{\sigma}_{P\!A} / 0.5~\mathrm{rad} ) \times P$.
Thus, we see that $\sigma_{q,u}$ is a function of three physical quantities: magnetic turbulence amplitude ($= \breve{\sigma}_{P\!A}$), dust alignment efficiency ($\propto P/A_\mathrm{G}$), and the extinction or gas column density ($\propto A_\mathrm{G} \propto \log N_\mathrm{H}$) as follows:
\begin{equation}
\sigma_{q,u} \simeq \frac{\breve{\sigma}_{P\!A}}{0.5~\mathrm{rad}} \times \frac{P}{A_\mathrm{G}} \times A_\mathrm{G}~.
\end{equation}
If we estimate $\sigma_{q,u}$, $P$, and $A_\mathrm{G}$ from observations, we can evaluate these physical quantities, including fluctuations in the position angle of the turbulent magnetic field in the POS.

The intrinsic values of $P$ and $A_\mathrm{G}$ for individual clouds can be found in Table \ref{tab:poleff}.
We can estimate the intrinsic $\sigma_{q,u}$ values specific to each cloud by subtracting the contributions from foreground polarization and observational uncertainties from the observed values (Equation (\ref{eq:sigma_intrinsic})).
This estimation assumes that the observed $\sigma_{q,u}$ is the squared sum of the intrinsic $\sigma_{q,u}$ and the contributions from foreground and observational uncertainties.
We measure $\sigma_{q,u\perp}$, which represents the $\sigma_{q,u}$ values in the direction perpendicular to each cloud's mean $q\mathchar`-u$ vector.
The estimated intrinsic $\sigma_{q,u\perp}$ values are presented in Table \ref{tab:sigma_tan}.
The $\sigma_{q,u\perp}$ values of the raw observed data and their observational uncertainties used for evaluating the intrinsic $\sigma_{q,u\perp}$ values are listed in Appendix \ref{sec:obs_values}.

\begin{table}[t]
\begin{center}
\caption{The turbulent magnetic field's angular amplitude.}
\label{tab:sigma_tan}
\begin{tabular}{crrr}
\hline
\hline
Cloud & \multicolumn{1}{c}{$\sigma_{q,u\perp}$} & \multicolumn{2}{c}{$\breve{\sigma}_{P\!A}$}\\
& \multicolumn{1}{c}{(\%)} & \multicolumn{1}{c}{(rad)} & \multicolumn{1}{c}{(deg)}\\
\hline
foreground & $0.19_{-0.03}^{+0.03}$ & $0.45_{-0.15}^{+0.22}$ & $25.9_{\;\,-8.4}^{+12.4}$\\
1.23 kpc & $0.14_{-0.14}^{+0.07}$ & $0.37_{-0.37}^{+0.29}$ & $21.0_{-21.0}^{+16.9}$\\
1.47 kpc & $0.43_{-0.06}^{+0.05}$ & $0.43_{-0.08}^{+0.10}$ & $24.4\;\,_{-4.8}^{+5.6}$\\
1.63 kpc & $0.26_{-0.14}^{+0.08}$ & $0.14_{-0.08}^{+0.06}$ & $8.1\;\,_{-4.4}^{+3.2}$\\
2.23 kpc & $0.29_{-0.13}^{+0.09}$ & $0.13_{-0.06}^{+0.06}$ & $7.3\;\,_{-3.4}^{+3.4}$\\
\hline
All & $0.64_{-0.02}^{+0.02}$ & $0.83_{-0.12}^{+0.16}$ & $47.8\;\,_{-6.9}^{+9.0}$\\
\hline
\multicolumn{4}{p{0.1\textwidth}}{\bf Notes.}\\
\multicolumn{4}{p{0.35\textwidth}}{The standard deviation of $q$ and $u$ is measured in the direction perpendicular to the mean $q\mathchar`-u$ vector ($\sigma_{q,u\perp}$) of each cloud,
and the amplitude of the turbulent magnetic field ($\breve{\sigma}_{P\!A}$) is estimated from $\sigma_{q,u\perp}$ and the polarization fraction ($P$)$^\mathrm{a}$, by referencing the theoretical function $\breve{\sigma}_{G(P\!A)}$ (Equation \ref{eq:norm2d}), considering the influence of the non-isotropic distribution of $\sigma_{q,u}$.}\\
\multicolumn{4}{p{0.35\textwidth}}{$^\mathrm{a}$ Intrinsic $P$ value of each cloud is taken from Table \ref{tab:poleff}.}\\
\end{tabular}
\end{center}
\end{table}

Similar to $\sigma_{q,u}$, the observed $\breve{\sigma}_{P\!A}$ is determined by the summation of contributions from multiple clouds along the LOS.
However, it should be noted that the addition of these contributions is not a simple linear sum of squares, as evident from the deviation of $\breve{\sigma}_{G(P\!A)}$ from the relation $\breve{\sigma}_{P\!A} \times P = \mathrm{const.}$ in Figures \ref{fig:s-p} and \ref{fig:norm2d}. Therefore, we estimate the intrinsic $\breve{\sigma}_{P\!A}$ values specific to individual clouds by referencing the $\sigma_{q,u}$ and $P$ values and the theoretical function $\breve{\sigma}_{G(P\!A)}$.

In the evaluation of $\breve{\sigma}_{G(P\!A)}$, we also take into account the non-isotropic distribution of $\sigma_{q,u}$, as described in Section \ref{sec:nonuniformity}. For each cloud, we determine the aspect ratio of the $\sigma_{q,u}$ distribution's major and minor axes, the rotation angle between the major axis and the average direction of the $q\mathchar`-u$ vectors, and the skewness of the $\sigma_{q,u}$ distribution both in radial and tangential directions. The obtained results are listed in Table \ref{tab:sigma_tan_isotropic} in Appendix \ref{seq:anisotropy}. We numerically calculate the deviation from the theoretical curve given by Equation (\ref{eq:norm2d}), taking into account the non-isotropic distribution of measured $\sigma_q$ and $\sigma_u$, and use the obtained theoretical curve to calculate $\breve{\sigma}_{P\!A}$. The values of $\breve{\sigma}_{P\!A}$ obtained considering the anisotropy of $\sigma_q$ and $\sigma_u$ vary within the range of estimation errors compared to the case where this consideration is omitted. The comparison of $\breve{\sigma}_{P\!A}$ estimation values with and without considering anisotropy is shown in Table \ref{tab:sigma_tan_isotropic}. Finally, we present the obtained values of $\breve{\sigma}_{P\!A}$, taking into account the anisotropy in the distribution of $\sigma_q$ and $\sigma_u$, in Table \ref{tab:sigma_tan}.

\subsubsection{Turbulent-to-Uniform Magnetic Field Intensity Ratio}
\label{sec:t-f ratio}

The estimated intrinsic $\breve{\sigma}_{P\!A}$ values shown in Table \ref{tab:sigma_tan} represent the amplitude of the turbulent magnetic field on the POS and can be used as indicators of the turbulent-to-uniform magnetic field intensity ratio $B_{\mathrm{turb}}/B_{\mathrm{unif}}$ when expressed in radians
\citep[e.g.,][]{1996ASPC...97..486Z,2008ApJ...679..537F,2021A&A...656A.118S}.

When $B_\text{unif}$ is not on the POS (i.e., for angles $i$ from the POS with $i>0^\circ$), the estimated value of $B_\text{turb}/B_\text{unif}$ from $\breve{\sigma}_{P\!A}$ can be overestimated depending on the value of $i$.
This is because the uniform magnetic field component projected onto the POS ($B_\text{unif, POS}$) has a dependence of $B_\text{unif, POS} = B_\text{unif}\cdot \cos (i)$ with respect to $i$, while the random component $B_\text{turb}$, if its distribution is isotropic, does not have a dependence on $i$. As a result, the observed angular dispersion $\breve{\sigma}_{P\!A}$ roughly increases proportionally to $[\cos (i)]^{-1}$ \citep[see, e.g.,][]{2008ApJ...679..537F,2013ApJ...777..112P,2018MNRAS.474.5122K,2019ApJ...887..159H}.

In the case where the large-scale magnetic field has an inclination of $i = 35^\circ$ corresponding to the pitch angle of the Sagittarius arm with respect to the POS ($\S$\ref{sec:each_cloud2}), it should be noted that the estimated $B_\text{turb}/B_\text{unif}$ values derived from the observed $\breve{\sigma}_{P\!A}$ shown in Table \ref{tab:sigma_tan} may be overestimated by a factor of 1.22 compared to the true value.

In the following, no correction for $i$ will be applied, and we will proceed with the discussion assuming $B_\text{turb}/B_\text{unif} \simeq \breve{\sigma}_{P\!A}$.

The obtained $B_{\mathrm{turb}}/B_{\mathrm{unif}}$ ratios range from 0.13 to 0.14 ($\simeq 7^\circ$ -- $8^\circ$) for the two more distant clouds, indicating that the magnetic fields associated with these clouds remain undisturbed by the random motions of the surrounding gas. In contrast, the closer three clouds exhibit $B_{\mathrm{turb}}/B_{\mathrm{unif}}$ ratios of approximately 0.37 to 0.45 ($\simeq 21^\circ$ -- $26^\circ$), suggesting a high degree of perturbation in their magnetic fields.

The observation covers a broader range for more distant clouds, potentially measuring a wider range of magnetic field orientations. However, the fact that we observe more ordered magnetic field orientations for more distant clouds is contrary to this expectation. This indicates that the observed difference in the degree of magnetic field perturbation is not due to the fact that the observation probes different spatial scales of the magnetic field for clouds at varying distances.

\begin{figure}[tp]
\centering
\includegraphics[width=\linewidth]{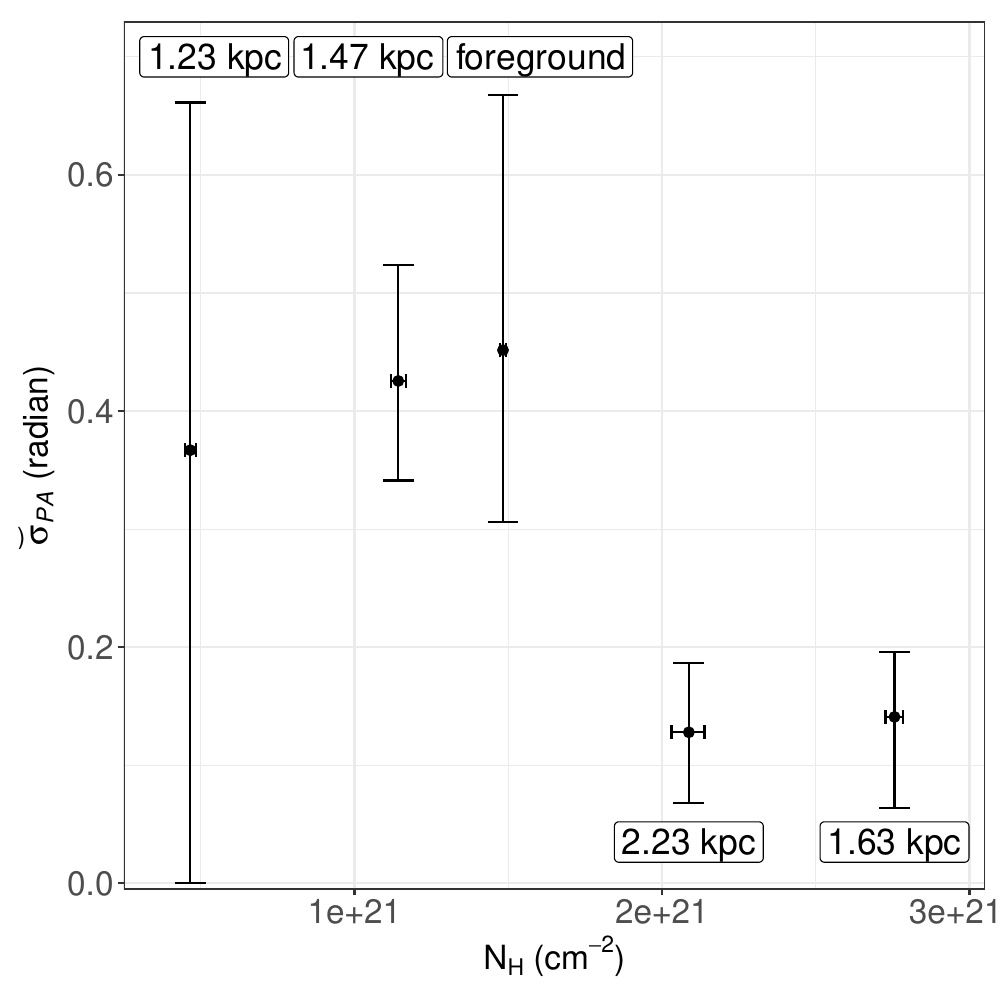}
\caption{
Correlation between the gas column density $N_\mathrm{H}$ and the turbulent magnetic field amplitude $\breve{\sigma}_{P\!A}$, which serves as an indicator of the turbulent-to-uniform magnetic field intensity ratio $B_{\mathrm{turb}}/B_{\mathrm{unif}}$.
The error bars of $\breve{\sigma}_{P\!A}$ indicate the 15.9\% and 84.1\% quantiles obtained from the 10,000 Monte Carlo simulations described in Section \ref{sec:each_cloud}.
}
\label{fig:nh_sigma}
\end{figure}

\subsubsection{Potential Correlation between \(\breve{\sigma}_{P\!A}\) and $N_\text{H}$}

Figure \ref{fig:nh_sigma} presents the dependence of $\breve{\sigma}_{P\!A}$ (which corresponds to $B_{\mathrm{turb}}/B_{\mathrm{unif}}$) on $N_\mathrm{H}$.
The two more distant clouds, where the magnetic fields remain undisturbed, exhibit relatively large column densities of $N_\mathrm{H} = 2.1$--$2.8 \times 10^{21}~\mathrm{cm}^{-2}$. In contrast, the three closer clouds, where the magnetic fields show a high degree of perturbation, have relatively small column densities of $N_\mathrm{H} = 0.5$--$1.5 \times 10^{21}~\mathrm{cm}^{-2}$ (Table \ref{tab:poleff}).
As a result, the correlation between $N_\mathrm{H}$ and $\breve{\sigma}_{P\!A}$ demonstrates a rough anti-correlation between the two variables.

Assuming equipartition between gas kinetic energy and the magnetic field energy, we can convert $B_\mathrm{turb}/B_\mathrm{unif}~(=\breve{\sigma}_{P\!A})$ to the magnetic field intensity.
Based on the discussions in \cite{2021A&A...647A.186S,2021A&A...656A.118S}, we estimate the magnetic field strength in the POS using the following equation:
\begin{equation}
\label{eq:ST}
B_\mathrm{POS} \approx \sqrt{2\pi\rho}\,\frac{\sigma_{v,N\!T}}{\sqrt{\sigma_\theta}},~
    \rho=\mu\,m_\mathrm{H}\,n_\mathrm{H},
\end{equation}
where $\sigma_{v,NT}$ is the gas non-thermal velocity dispersion, $\sigma_\theta$ is the position angle dispersion of the polarization due to the turbulent magnetic field ($=\breve{\sigma}_{P\!A}$), $\rho$ is the gas density, $\mu$ is the average particle mass (including hydrogen and helium), $m_\mathrm{H}$ is the hydrogen atomic mass, and $n_\mathrm{H}$ is the gas number density.

Since the gas is primarily atomic (Section \ref{sec:results}), we adopt $\mu=1.4$ \citep{2022arXiv220311179P}.
As a result, we obtain the following values for the magnetic field strength in each cloud.
\begin{eqnarray}
\left[
12.1_{-2.2}^{+2.6},\,13.5_{-3.4}^{+\infty},\,12.5_{-1.2}^{+1.5},\,21.7_{\;\,-3.3}^{+10.6},\,22.8_{-3.9}^{+8.4}
\right] \nonumber\\
\times \left( \frac{n_\mathrm{H}}{10^2\,\mathrm{cm}^{-3}} \right)^{1/2} \left( \frac{\Delta V_\mathrm{FWHM}}{5\,\mathrm{km\,s}^{-1}} \right) \mu \mathrm{G}, \nonumber
\end{eqnarray}
for the foreground, 1.23 kpc, 1.47 kpc, 1.63 kpc, and 2.23 kpc clouds, respectively.
Here we normalize the results by the reference values of the gas number density $n_\mathrm{H} = 10^2\,\mathrm{cm}^{-3}$ and the full width at half maximum value of gas velocity $\Delta V_\mathrm{FWHM} = 5\,\mathrm{km\,s}^{-1}$, since no estimated values are available (Section \ref{sec:each_cloud2}).
The assumption of $n_\mathrm{H} = 10^2\,\mathrm{cm}^{-3}$ is equivalent to assuming the LOS thickness of about 1.5 -- 8.9 pc for the five clouds in the LOS, whose column densities are 0.47 -- $2.76\times 10^{21}\,\mathrm{cm}^{-2}$ (Table \ref{tab:poleff}).
We apply $\Delta V_\mathrm{FWHM} = 5\,\mathrm{km\,s}^{-1}$ as a proxy for the HI gas velocity dispersion around CO cloud cores since \cite{2015ApJS..216...18N} observed the Orion A and Orion B molecular clouds in $^{12}\mathrm{CO}(J=2\mathchar`-\mathchar`-1)$ emission line and estimated that the linewidths ($\simeq \Delta V_\mathrm{FWHM}$) are generally 2 -- 5 $\mathrm{km~s}^{-1}$.
If we adopt a velocity dispersion of $\Delta V_\mathrm{FWHM} = 2\,\mathrm{km\,s}^{-1}$, it should be noted that the estimated magnetic field strength will proportionally decrease. On the other hand, if the uniform magnetic field is inclined $i = 35^\circ$ from the POS, the estimated magnetic field strength will be approximately 10\% larger (by a factor of $\sqrt{1.22}$, see Section \ref{sec:t-f ratio}).

\cite{1980BAAS...12..860H} measured the global magnetic field strength associated with the H{\small I} gas in the Sagittarius arm by the Zeeman splitting of 21 cm H{\small I} emission line in the tangential direction ($l = 51^\circ,~b=0^\circ$) of the Sagittarius arm.
They obtained a value of $\sim 16\,\mu$G.
Our estimated magnetic field strengths for the clouds in the Sagittarius arm are roughly consistent with their estimation, although the magnetic field strengths estimate given here will change depending on the gas velocity dispersion and density.

Using OH Zeeman effect measurements, \cite{2012ARA&A..50...29C} found that the magnetic field strength in low column density interstellar clouds is typically $\sim 10\,\mu\mathrm{G}$, which makes these clouds magnetically subcritical.
In contrast, clouds with higher column densities are often magnetically supercritical; the gas contracts gravitationally and drags the magnetic field lines inward.
The column density threshold at which the cloud transition from subcritical to supercritical is $N_\mathrm{H} \sim 2 \times 10^{21}~\mathrm{cm}^{-2}$ \citep[also see][for a review]{2022arXiv220311179P}.

The column density of each cloud we observe corresponds to this threshold value, as indicated in Figure \ref{fig:nh_sigma}.
Clouds with higher column densities tend to exhibit more ordered magnetic fields (Figure \ref{fig:nh_sigma}), which can be indicative of stronger magnetic field strengths ($\sim 20\,\mu$G). The higher degree of orderliness may be the result of gas contraction, indicating that these clouds may be magnetically supercritical and the gravitational forces within the cloud are comparable to or greater than the magnetic pressure and the thermal pressure. The presence of a stronger and ordered magnetic field can have significant implications for the dynamics and evolution of the cloud, influencing processes such as star formation and gas kinematics.
Thus, our method of separately estimating the magnetic field turbulence and polarization efficiency of each cloud along the LOS can be a promising means of observing interstellar clouds that are magnetically near-critical.

As shown in Figures \ref{fig:Sagittarius_qu_cloud} and \ref{fig:Sagittarius_qu_diff}, we have successfully detected multiple flips in $P\!A$ along the LOS in the Sagittarius arm.
These flips may be due to the local deformation of the magnetic field around individual clouds.
However, each cloud's magnetic field is smooth within the observed region (Figures \ref{fig:Sagittarius_qu_cloud}, \ref{fig:Sagittarius_qu_diff}).
The physical size of the observed region is 3.8 pc $\times$ 6.1 pc for the 1.23 kpc cloud and 6.8 pc $\times$ 11 pc for the 2.23 kpc cloud.
Thus, the results indicate the possibility of a global magnetic field flipping on a scale sufficiently larger than the observed area and possibly even larger than the size of individual clouds.
This possibility should be investigated by observing several neighboring regions.

\section{Summary} \label{sec:summary}

We completed an $R_\mathrm{C}$-band polarimetric survey around $l = 14^\circ\negthinspace.15,~ b=-1^\circ\negthinspace.47$ in a direction that threads the Sagittarius spiral arm using HONIR, an imaging polarimeter on the Kanata Telescope, Hiroshima University.
We selected a region where a large number of {\it Gaia} stars were measured with sufficient precision.
We found that the polarization position angles ($=$ position angles of the magnetic field projected onto the POS) in the LOS vary significantly at each of four locations at distances of 1.23 kpc, 1.47 kpc, 1.63 kpc, and 2.23 kpc.
Based on these data, we found four isolated clouds at these locations in the LOS and a foreground cloud at $d < 200$ pc, which is possibly an outskirt of the Aquila Rift, for a total of five clouds, producing the observed polarization.

The column density of each cloud is $\lesssim 2.8\times 10^{21}~\mathrm{cm}^{-2}$.
No corresponding CO molecular clouds are found in the literature, suggesting that these clouds are primarily atomic and may be the surroundings of denser molecular clouds.
Thanks to the distinct change in the magnetic fields’ position angles, we have detected tenuous dust clouds along the LOS with high sensitivity.

We successfully extracted the magnetic field characteristics of each cloud by differencing the polarimetry data and the {\it Gaia} stellar extinction data by distance.
Individual clouds' estimated magnetic field structure is smooth within the $17\arcmin \times 10\arcmin$ observed region.
The scale length of the structure is thus expected to be $\gtrsim 10'$, corresponding to $\gtrsim 10$ pc in physical scales.

Individual clouds' magnetic field position angles are $134.5_{-2.8}^{+2.8}$ deg, $46.1_{-4.8}^{+4.7}$ deg, $58.1_{-2.8}^{+2.8}$ deg, $150.2_{-1.4}^{+1.4}$ deg, and $40.3_{-1.2}^{+1.2}$ deg in East of North of the Galactic coordinates for the clouds in increasing order of distance.
The position angles are significantly offset from the direction of the Galactic plane, with deviations of approximately $\pm (30^\circ \mathchar`-\mathchar`- 60^\circ)$, in contrast to the current understanding that the large-scale magnetic field in the Galactic disk is parallel to the Galactic plane.

The polarization efficiency of each dust cloud is $P / A_\mathrm{G} = 0.4\%\,\mathrm{mag}^{-1}$ for the foreground cloud and 1.0 -- $1.4\%\,\mathrm{mag}^{-1}$ for each dust cloud in the Sagittarius arm.
These values are comparable to or lower than those of Taurus and Perseus \citep[$1.5\%\,\mathrm{mag}^{-1}$;][]{2021ApJ...914..122D}.
Besides the angular offset from the Galactic plane mentioned above, the magnetic field of individual clouds may be inclined to the POS at different angles, causing slightly lower polarization efficiency of clouds and their variation.

The turbulent amplitude of the magnetic field associated with each cloud ($\breve{\sigma}_{P\!A}$), which can be used as indicators of the turbulent-to-uniform magnetic field intensity ratio $B_{\mathrm{turb}}/B_{\mathrm{unif}}$, is weakly correlated with the column density of each dust cloud, ranging from $\breve{\sigma}_{P\!A} = 21^\circ\negthinspace.0$ -- $25^\circ\negthinspace.9$ ($B_{\mathrm{turb}}/B_{\mathrm{unif}} = 0.37$ -- 0.45) for three clouds with relatively low column density $N_\mathrm{H} = 0.47$ -- $1.48 \times 10^{21}\,\mathrm{cm}^{-2}$ and $\breve{\sigma}_{P\!A} = 7^\circ\negthinspace.3$ -- $8^\circ\negthinspace.1$ ($B_{\mathrm{turb}}/B_{\mathrm{unif}} = 0.13$ -- 0.14) for two clouds with relatively high column density $N_\mathrm{H} = 2.09$ -- $2.76 \times 10^{21}\,\mathrm{cm}^{-2}$.
Assuming the general values $n_\mathrm{H}=10^2\,\mathrm{cm}^{-3}$ and $\Delta V _\mathrm{FWHM} = 5\,\mathrm{km}\,\mathrm{s}^{-1}$ as the gas density and gas velocity dispersion, we estimated the magnetic field strength 12 -- $13\,\mu$G for the low column density clouds and $\sim 20\,\mu$G for the high column density ones.

Our observations show an anti-correlation between polarization angular dispersion $\sigma_{P\!A}$ and polarization fraction $P$ that was found by Planck observations \citep{2020A&A...641A..12P}.
We showed that this anti-correlation can be obtained from the shift of data points in the $q\mathchar`-u$ plane while keeping $\sigma_{q,u}$ in each region constant due to geometrical depolarization.
The magnetic field structure of each region we observed is smooth, even at scales below Planck's spatial resolution.
Therefore, the difference between our observations and Planck's is likely due to the difference in distances probed instead of differences in beam sizes.

As demonstrated above, by combining optical polarimetry data with {\it Gaia} catalog distances and interstellar extinction, we could estimate each cloud's magnetic turbulence and polarization efficiency along the LOS separately.
We argue that this method is a functional tool for investigating the turbulent nature of the magnetic field at the periphery of interstellar molecular clouds in 3D.

\begin{acknowledgments}
The authors are grateful to an anonymous referee, who provided thorough and thoughtful suggestions for improving various aspects of the paper.
This work has made use of data from the European Space Agency (ESA) mission {\it Gaia} (\url{https://www.cosmos.esa.int/gaia}), processed by the {\it Gaia} Data Processing and Analysis Consortium (DPAC, \url{https://www.cosmos.esa.int/web/gaia/dpac/consortium}).
Funding for the DPAC has been provided by national institutions, in particular the institutions participating in the {\it Gaia} Multilateral Agreement.
This research has been supported by JSPS KAKENHI grants 25247016, 18H01250, 18H03720, 20K03276, and 20K04013.
M.Tahani is supported by the Banting Fellowship (Natural Sciences and Engineering Research Council Canada) hosted at Stanford University and the Kavli Institute for Particle Astrophysics and Cosmology (KIPAC) Fellowship.
CVR thanks the Brazilian Conselho Nacional de Desenvolvimento Científico e Tecnológico - CNPq (Proc: 310930/2021-9).
AMM’s work and optical/NIR polarimetry at IAG have been supported over the years by several grants from the São Paulo state funding agency FAPESP, especially 01/12589-1 and 10/19694-4. AMM has also been partially supported by the Brazilian agency CNPq  (grant 310506/2015-8). AMM graduate students have received grants over the years from the Brazilian agency CAPES.
\end{acknowledgments}


\vspace{5mm}
\facilities{Kanata:1.5m (HONIR), {\it Gaia}, Planck}


\software{
strucchange \citep{strucchange,breakpoints},
astropy \citep{2013A&A...558A..33A,2018AJ....156..123A,2022ApJ...935..167A},  
Source Extractor \citep{1996A&AS..117..393B}
}



\appendix

\section{Coordinate conversion of the normalized Stokes parameters {\it q} and {\it u}}
\label{sec:coord_conv}

The measured normalized Stokes parameters, $q$ and $u$, are defined in equatorial coordinates. We convert these values into Galactic coordinates, $q_\mathrm{Gal}$ and $u_\mathrm{Gal}$, to align the polarization measurements with the Galactic coordinate system for the discussion in this paper.

To perform the conversion, we first evaluate the position angle offset between the Galactic and equatorial coordinates ($\phi$) at each stellar position using the following equations based on spherical trigonometry:
\begin{equation}
\begin{aligned}
P\!A_\mathrm{Gal} & = P\!A_\mathrm{eq} + \phi,\\
%\phi & = \arctan\left(\frac{Y}{X}\right), \\
\phi & = \arctantwo\left(Y, X \right), \\
\text{where,} \quad Y & = \cos(\delta_\mathrm{NGP}) \times \sin(\alpha - \alpha_\mathrm{NGP}),\\
X & = \sin(\delta_\mathrm{NGP}) \times \cos(\delta) - \cos(\delta_\mathrm{NGP}) \times \sin(\delta) \times \cos(\alpha - \alpha_\mathrm{NGP}).
\end{aligned}
\end{equation}
Here, $(\alpha ,~ \delta)$ represents the equatorial coordinate (J2000) position of the source, and $(\alpha_\mathrm{NGP} ,~ \delta_\mathrm{NGP})$ represents the equatorial coordinate (J2000) position of the north Galactic pole \citep{2000asqu.book.....C}:

\begin{equation}
\begin{aligned}
\alpha_\mathrm{NGP} & = 3.366033 \;\; \mathrm{radians}, \\
\delta_\mathrm{NGP} & = 0.4734773 \, \mathrm{radians}.
\end{aligned}
\end{equation}

Based on $P\!A_\mathrm{Gal}$ and the length of the $q\mathchar`-u$ pseudo-vector $\sqrt{q^2 + u^2}$, we estimate the normalized Stokes parameters in Galactic coordinates:

\begin{equation}
\begin{aligned}
q_\mathrm{Gal} &= \sqrt{q^2 + u^2} \cdot \cos \left(2 \cdot P\!A_\mathrm{Gal}\right), \\
u_\mathrm{Gal} &= \sqrt{q^2 + u^2} \cdot \sin \left(2 \cdot P\!A_\mathrm{Gal}\right).    
\end{aligned}
\end{equation}

\section{Data list}
\label{sec:data_list}

The data used for the analysis (Section \ref{sec:GaiaID}) are the {\it Gaia} catalog stars and their optical polarization data, which are shown in Table \ref{tab:data_list}. The data consist of 313 stars within the observed FOV. Among them, optical polarimetry data is available for 184 stars, and $A_\mathrm{G}$ values are available for 259 stars. It should be noted that the polarization values, $q_\mathrm{Gal}$ and $u_\mathrm{Gal}$, are given in Galactic coordinates (refer to Section \ref{sec:calibration} and Appendix \ref{sec:coord_conv}). These values represent the observed apparent values and are not corrected for foreground contributions. The apparent values of the polarization position angle and its uncertainty, $P\!A$ and $\delta P\!A$, can be calculated as follows:
\begin{equation}
\begin{aligned}
P\!A &= 0.5 \times \arctan\left(\frac{u_\mathrm{Gal}}{q_\mathrm{Gal}}\right), \\
\delta P\!A &= \frac{\sqrt{q_\mathrm{Gal}^2 \cdot \delta u_\mathrm{Gal}^2 + u_\mathrm{Gal}^2 \cdot \delta q_\mathrm{Gal}^2}}{2\times (q_\mathrm{Gal}^2+u_\mathrm{Gal}^2)}.
\end{aligned}
\end{equation}

We calculate the apparent value of the polarization degree, $P$, using the observed values of $q_\mathrm{Gal}$ and $u_\mathrm{Gal}$, along with their uncertainties, $\delta q_\mathrm{Gal}$ and $\delta u_\mathrm{Gal}$.
Taking into account their positive bias within the squared sum of $q_\mathrm{Gal}$ and $u_\mathrm{Gal}$ values, noteworthy in low signal-to-noise polarization signals (e.g., $q/\delta q < 3$ and $u/\delta u < 3$; \citealp{1974ApJ...194..249W,2006PASP..118.1340V}), we emphasize the need for cautious utilization of such signals.
The computation of $P$, while considering its positive bias and the associated uncertainty, $\delta P$, is performed following the methodology outlined in \cite{1974ApJ...194..249W,2006PASP..118.1340V}:
\begin{equation}
\begin{aligned}
\delta P &= \sqrt{\frac{q_\mathrm{Gal}^2 \cdot \delta q_\mathrm{Gal}^2+u_\mathrm{Gal}^2 \cdot \delta u_\mathrm{Gal}^2}{q_\mathrm{Gal}^2+u_\mathrm{Gal}^2}}, \\
P &= \sqrt{q_\mathrm{Gal}^2+u_\mathrm{Gal}^2 - \delta P^2}.
\label{eq:debias}
\end{aligned}
\end{equation}
These values are also presented in Table \ref{tab:data_list}.

\begin{table*}
\begin{center}
\caption{Data Identification}
\label{tab:data_list}
\begin{splittabular}{lrrrrcccBrrrrrrrrccc}
\hline
\hline
Gaia ID$^\mathrm{a}$ & \multicolumn{1}{c}{R.A.$^\mathrm{b}$} & \multicolumn{1}{c}{Dec.$^\mathrm{b}$} & \multicolumn{1}{c}{$l$} & \multicolumn{1}{c}{$b$} & r\_med\_geo$^\mathrm{c}$ & r\_lo\_geo$^\mathrm{c}$ & r\_hi\_geo$^\mathrm{c}$ & \multicolumn{1}{c}{$q_\mathrm{Gal}$} & \multicolumn{1}{c}{$\delta q_\mathrm{Gal}$} & \multicolumn{1}{c}{$u_\mathrm{Gal}$} & \multicolumn{1}{c}{$\delta u_\mathrm{Gal}$} & \multicolumn{1}{c}{$P^\mathrm{d}$} & \multicolumn{1}{c}{$\delta P$} & \multicolumn{1}{c}{$P\!A_\mathrm{Gal}$} & \multicolumn{1}{c}{$\delta P\!A_\mathrm{Gal}$} & ag\_gspphot$^\mathrm{e}$ & ag\_gspphot\_lower$^\mathrm{e}$ & ag\_gspphot\_upper$^\mathrm{e}$\\
& \multicolumn{1}{c}{(deg)} & \multicolumn{1}{c}{(deg)} & \multicolumn{1}{c}{(deg)} & \multicolumn{1}{c}{(deg)} & \multicolumn{1}{c}{(pc)} & \multicolumn{1}{c}{(pc)} & \multicolumn{1}{c}{(pc)} & & & & & \multicolumn{1}{c}{(\%)} & \multicolumn{1}{c}{(\%)} & \multicolumn{1}{c}{(deg)} & \multicolumn{1}{c}{(deg)} & \multicolumn{1}{c}{(mag)} & \multicolumn{1}{c}{(mag)} & \multicolumn{1}{c}{(mag)}\\
\hline
4096570661993470336 & 275.3631 & -17.34921 & 14.13137 & -1.440729 & 172.7874 & 170.6747 & 175.1837 &  &  &  &  &  &  &  &  & 0.7675 & 0.7522 & 0.7822\\
4096570043496229632 & 275.4721 & -17.35551 & 14.17472 & -1.535573 & 179.5864 & 178.6372 & 180.5435 & 0.0027 & 0.0015 & -0.0046 & 0.0020 & 0.47 & 0.26 & 150.31 & 7.11 & 0.9327 & 0.9260 & 0.9377\\
4096549530732278528 & 275.2764 & -17.39164 & 14.05501 & -1.387679 & 238.9407 & 228.1535 & 247.4192 &  &  &  &  &  &  &  &  & 0.4863 & 0.4419 & 0.5229\\
4096550256663944192 & 275.3030 & -17.35526 & 14.09908 & -1.393004 & 250.0807 & 248.8838 & 250.9728 & 0.0016 & 0.0009 & -0.0016 & 0.0013 & 0.15 & 0.17 & 157.39 & 11.47 & 0.8379 & 0.8000 & 0.9082\\
4096546850671183744 & 275.3928 & -17.39369 & 14.10546 & -1.486728 & 325.4934 & 323.6708 & 327.0741 &  &  &  &  &  &  &  &  & 0.5252 & 0.5203 & 0.5444\\
4096574063597660032 & 275.3781 & -17.29108 & 14.18945 & -1.426083 & 358.5222 & 355.7014 & 361.4063 & 0.0020 & 0.0008 & -0.0042 & 0.0006 & 0.46 & 0.06 & 147.48 & 5.48 & 0.6387 & 0.6328 & 0.6493\\
4096546919392208512 & 275.3832 & -17.38007 & 14.11315 & -1.472199 & 386.3853 & 380.1137 & 391.5740 &  &  &  &  &  &  &  &  & 0.8166 & 0.8132 & 0.8223\\
4096571005567251968 & 275.4739 & -17.32008 & 14.20684 & -1.520485 & 387.6165 & 384.3686 & 391.0230 & -0.0010 & 0.0005 & -0.0016 & 0.0006 & 0.18 & 0.05 & 118.38 & 10.17 & 0.3669 & 0.3251 & 0.3974\\
4096574097945622016 & 275.3504 & -17.29481 & 14.17370 & -1.404466 & 450.7833 & 442.4085 & 459.0101 & 0.0003 & 0.0024 & -0.0047 & 0.0018 & 0.43 & 0.20 & 136.55 & 16.29 & 0.5510 & 0.5393 & 0.5622\\
4096571284825275008 & 275.4531 & -17.29941 & 14.21573 & -1.493203 & 479.7004 & 472.8917 & 486.8683 &  &  &  &  &  &  &  &  & 0.7815 & 0.7771 & 0.7863\\
\hline
\end{splittabular}
\end{center}
{\bf Notes.}\\
$^\mathrm{a}$ {\it Gaia} source ID in DR 3.\\
$^\mathrm{b}$ {\it Gaia} DR 3 coordinate positions. The reference epoch is 2016.\\
$^\mathrm{c}$ Distance to the stars (r\_med\_geo) and their 16th and 84th percentiles (r\_lo\_geo and r\_hi\_geo) estimated by \cite{2021AJ....161..147B}.\\
$^\mathrm{d}$ The debiased values of \(P\) as determined by Equation (\ref{eq:debias}) are shown. Due to the debiasing process, it is possible for the \(P^2\) value in Equation (\ref{eq:debias}) to be negative. In the online data files, \(P\) values for data points with negative \(P^2\) values are indicated as '$<0$'.\\
$^\mathrm{e}$ Extinction in G band ($A_\mathrm{G}$) values (ag\_gspphot) and  their 16th and 84th percentiles (ag\_gspphot\_lower and ag\_gspphot\_upper) taken from the {\it Gaia} DR3 catalog \citep{2022arXiv220800211G}.\\
(This table is available in its entirety in machine-readable form.
A portion is shown here for guidance regarding its form and content.)
\end{table*}

\section{Breakpoint analysis}
\label{sec:breakpoint}

A breakpoint analysis is a statistical method for estimating the location of abrupt changes in a data sequence.
We assume the number of breakpoints in a data sequence a priori, perform least-squares fits of the data, and obtain the most likelihood estimations (MLE) of where those breakpoints occur.
We then repeat the fit with different numbers of breakpoints and compare the goodness-of-fit values using the Bayesian information criterion \citep{akaike1978new, 1978AnSta...6..46}, to get the most likely numbers of breakpoints and their positions.

We assume that $q_\mathrm{Gal}$ and $u_\mathrm{Gal}$ are constant as a function of distance, which corresponds to the assumption that the observed polarization is caused by 2D sheet(s) of ISM at specific distance(s).
In addition, we assume the $q_\mathrm{Gal}$ and $u_\mathrm{Gal}$ distributions have a certain number of step-wise changes (i.e., breakpoints), which correspond to the positions of the 2D sheets.

We perform the breakpoint analysis for distance dependence of $q_\mathrm{Gal}$ and $u_\mathrm{Gal}$ shown in Figure \ref{fig:Sagittarius_qu_dist} by using the R library `strucchange' \citep{strucchange,breakpoints}.
To make a stable detection of the 2D change of $q_\mathrm{Gal}$ and $u_\mathrm{Gal}$ on the $q\mathchar`-u$ plane, we perform breakpoint analysis for inner product of each $q\mathchar`-u$ vector with the unit vector of phase angle $\theta$.
We then repeat the breakpoint analysis by changing $\theta$ from $0^\circ$ to $360^\circ$ by every $5^\circ$ and adopt breakpoints detected at a continuous θ of more than $90^\circ$.

\section{Galactic surface density of dust clouds}
\label{sec:surface_density}

We estimate the surface density of dust clouds in the Galaxy by calculating their all-sky 3D distribution. Breakpoint analyses are conducted on the celestial plane at intervals of $6\arcmin\negthinspace.87$, using stars with 'geometric' distances \citep[][with uncertainties $\leqslant 20$\%]{2021AJ....161..147B} and estimations of $A_\mathrm{G}$ (DR3 catalog). The spatial resolution of our estimation is set at $15\arcmin$. Along each line of sight, the distance to the dust clouds is estimated. Median $A_\mathrm{G}$ values between breakpoints are computed and differentiated to assess the increase in $A_\mathrm{G}$ value at each breakpoint position, representing the column density of dust clouds at those positions. To convert $A_\mathrm{G}$ to hydrogen column density, we assume $N_\mathrm{H} = A_\mathrm{G} \cdot 2.21 \times 10^{21}/0.789$ \citep{2009MNRAS.400.2050G,2019ApJ...877..116W}. Finally, we estimate the surface density of dust clouds by integrating the column density within a range of $\pm 100$ pc from the Galactic plane, using 20 pc $\times$ 20 pc bins on the Galactic plane.

\section{Average observed values for each distance range}
\label{sec:obs_values}

To estimate the intrinsic physical parameters of the magnetic field associated with each dust cloud, we follow a two-step process. First, we calculate the average observed values for each distance range that corresponds to each dust cloud along the LOS. These average values represent the measured properties of the cloud. Then, to obtain the intrinsic values, we subtract the average value of the immediately preceding cloud from the average value of the specific cloud. This difference provides an estimation of the intrinsic physical parameters specific to each dust cloud, which helps us understand the properties of the magnetic field associated with each cloud in a more accurate and meaningful way.

Table \ref{tab:poleff_obs} displays the average polarization fraction ($P$), average $A_\mathrm{G}$, and the average $N_\mathrm{H}$ values. The $N_\mathrm{H}$ values are derived from $A_\mathrm{G}$ using the conversion $N_\mathrm{H} = A_\mathrm{G} \cdot 2.21 \times 10^{21}/0.789$ \citep{2009MNRAS.400.2050G,2019ApJ...877..116W}. These average observed values serve as the basis for estimating the intrinsic values, which are presented in Table \ref{tab:poleff}.
Table \ref{tab:sigma_tan_obs} displays the standard deviation of $q$ and $u$ measured in the direction perpendicular to the mean $q\mathchar`-u$ vector ($\sigma_{q,u\perp}$) and their observational uncertainties within each distance range.
These values serve as the basis for estimating the intrinsic $\sigma_{q,u\perp}$ values, which are presented in Table \ref{tab:sigma_tan}.

\begin{table*}[t]
\begin{center}
\caption{Average polarization fraction, interstellar extinction and column density values within each distance range.}
\label{tab:poleff_obs}
\begin{tabular}{cccrr}
\hline
\hline
Distance range & Cloud & Polarization fraction ($P$) & \multicolumn{1}{c}{$A_\mathrm{G}$} & \multicolumn{1}{c}{$N_\mathrm{H}^\mathrm{~a}$}\\
(kpc) & & \multicolumn{1}{c}{(\%)} & \multicolumn{1}{c}{(mag)} & \multicolumn{1}{c}{($10^{21}\mathrm{cm}^{-2}$)}\\
\hline
~~~~~~-- 1.23 & foreground & $0.22_{-0.02}^{+0.02}$ & $0.53_{-0.00}^{+0.00}$ & $1.48_{-0.01}^{+0.01}$ \\
1.23 -- 1.47 & 1.23 kpc & $0.04_{-0.02}^{+0.02}$ & $0.70_{-0.01}^{+0.01}$ & $1.95_{-0.01}^{+0.02}$ \\
1.47 -- 1.63 & 1.47 kpc & $0.53_{-0.04}^{+0.04}$ & $1.10_{-0.01}^{+0.01}$ & $3.09_{-0.02}^{+0.02}$ \\
1.63 -- 2.23 & 1.63 kpc & $0.46_{-0.02}^{+0.02}$ & $2.09_{-0.01}^{+0.01}$ & $5.84_{-0.02}^{+0.02}$ \\
2.23 --~~~~~~ & 2.23 kpc & $0.76_{-0.03}^{+0.03}$ & $2.83_{-0.02}^{+0.02}$ & $7.94_{-0.05}^{+0.05}$ \\
\hline
\multicolumn{5}{p{0.25\textwidth}}{\bf Notes.}\\
\multicolumn{5}{p{0.60\textwidth}}{$^\mathrm{a}$ $N_\mathrm{H} = A_\mathrm{G} \cdot 2.21 \times 10^{21}/0.789$ is assumed \citep{2009MNRAS.400.2050G,2019ApJ...877..116W}.}\\
\end{tabular}
\end{center}
\end{table*}

\begin{table}[t]
\begin{center}
\caption{Standard deviation and uncertainties in observed $q$ and $u$.}
\label{tab:sigma_tan_obs}
\begin{tabular}{ccrr}
\hline
\hline
Distance range & Cloud & \multicolumn{2}{c}{$\sigma_{q,u\perp}$} \\
& & \multicolumn{1}{c}{Observed} & \multicolumn{1}{c}{Uncertainty} \\
& & \multicolumn{1}{c}{(\%)} & \multicolumn{1}{c}{(\%)} \\
\hline
~~~~~~-- 1.23 & foreground & $0.24_{-0.02}^{+0.03}$ & $0.14_{-0.02}^{+0.02}$ \\
1.23 -- 1.47 & 1.23 kpc & $0.26_{-0.04}^{+0.04}$ & $0.16_{-0.03}^{+0.03}$ \\
1.47 -- 1.63 & 1.47 kpc & $0.52_{-0.04}^{+0.04}$ & $0.20_{-0.03}^{+0.04}$ \\
1.63 -- 2.23 & 1.63 kpc & $0.58_{-0.02}^{+0.02}$ & $0.18_{-0.02}^{+0.02}$ \\
2.23 --~~~~~~ & 2.23 kpc & $0.63_{-0.05}^{+0.06}$ & $0.17_{-0.03}^{+0.03}$ \\
\hline
\multicolumn{4}{p{0.25\textwidth}}{\bf Notes.}\\
\multicolumn{4}{p{0.42\textwidth}}{The standard deviation of $q$ and $u$ measured in the direction perpendicular to the mean $q\mathchar`-u$ vector ($\sigma_{q,u\perp}$) within each distance range, along with their associated observational uncertainties, is presented.}\\
\end{tabular}
\end{center}
\end{table}

\section{Anisotropy Observed in $\sigma_q$ and $\sigma_u$ Distributions}
\label{seq:anisotropy}

As described in Section \ref{sec:nonuniformity}, the dependence of $\breve{\sigma}_{P\!A}$ on $P/\sigma_{q,u}$ deviates from the theoretical value $\breve{\sigma}_{G(P\!A)}$ shown in Equation (\ref{eq:norm2d}), when $\sigma_q$ and $\sigma_u$ are not isotropically distributed with respect to their phase angles in the q-u plane. In addition to the phase angle dependence of the observed $\sigma_q$ and $\sigma_u$ values, the non-Gaussian distribution of observed $q$ and $u$ values on the $q\text{-}u$ plane (a distribution with non-zero skewness) produces deviations from Equation (\ref{eq:norm2d}). By numerically determining the dependence of $\breve{\sigma}_{P\!A}$ on $P/\sigma_{q,u}$, considering these two effects, we estimate the most likely intrinsic $\breve{\sigma}_{P\!A}$ value for each cloud and compare it with the value obtained without considering the anisotropy effect.

We approximate the distribution of $\sigma_{q}$ and $\sigma_{u}$ as an ellipse and estimate the aspect ratio of the major and minor axes, along with the rotation angle (the angle between the major axis of the elliptical approximation and the $q\text{-}u$ mean vector) of the ellipse. Additionally, we estimate the skewness of the $q$ and $u$ data distribution on the $q\text{-}u$ plane, measured in the directions parallel (radial) and perpendicular (tangential) to the $q\text{-}u$ mean vector. By accounting for these measures of ellipticity and skewness, we then numerically evaluate the $\breve{\sigma}_{P\!A}$ value based on the $P/\sigma_{q,u}$ value. We perform Monte Carlo simulations for 10,000 repetitions (as detailed in Section \ref{sec:each_cloud}), obtaining the median as the maximum likelihood estimate of $\breve{\sigma}_{P\!A}$, and determining the 15.9\% and 84.1\% quantiles as the negative and positive errors, respectively.

The aspect ratio and rotation angle of the obtained ellipses, along with the skewness of the $q$ and $u$ data distributions (both radial and tangential), as well as the resulting $\breve{\sigma}_{P\!A}$ values, are presented in Table \ref{tab:sigma_tan_isotropic}. The table displays two estimates of $\breve{\sigma}_{P\!A}$: one accounting for the non-isotropic distribution of $\sigma_q$ and $\sigma_u$ (`non-isotropic' in Table \ref{tab:sigma_tan_isotropic}), and the other assuming an isotropic distribution (`isotropic' in Table \ref{tab:sigma_tan_isotropic}). As evident from Table \ref{tab:sigma_tan_isotropic}, the two values of $\breve{\sigma}_{P\!A}$ are consistent within their respective error ranges, indicating that the influence of the non-isotropic distribution is small.

In Table \ref{tab:sigma_tan}, we present the $\breve{\sigma}_{P\!A}$ values considering the non-isotropic distribution, and from Section \ref{sec:nonuniformity} onward, we continue the discussion based on these $\breve{\sigma}_{P\!A}$ values that account for the non-isotropic distribution.

\begin{table}[t]
\begin{center}
\caption{Differences in estimated \(\breve{\sigma}_{P\!A}\) due to consideration of non-isotropic distributions of \(\sigma_q\) and \(\sigma_u\)}
\label{tab:sigma_tan_isotropic}
\begin{tabular}{crrrrrr}
\hline
\hline
Cloud & \multicolumn{1}{c}{Aspect Ratio$^\text{a}$} & \multicolumn{1}{c}{Rotation Angle$^\text{b}$} & \multicolumn{2}{c}{Skewness} & \multicolumn{1}{c}{$\breve{\sigma}_{P\!A}(\text{isotropic})^\text{c}$} & \multicolumn{1}{c}{$\breve{\sigma}_{P\!A}(\text{non-isotropic})^\text{d}$}\\
& & \multicolumn{1}{c}{(deg)} & \multicolumn{1}{c}{radial} & \multicolumn{1}{c}{tangengial} & \multicolumn{1}{c}{(deg)} & \multicolumn{1}{c}{(deg)}\\
\hline
foreground & $1.38_{-0.19}^{+0.24}$ & $45.8_{-35.2}^{+31.7}$ & $0.17_{-0.47}^{+0.47}$ & $0.01_{-0.42}^{+0.43}$ & $27.5\;\,_{-4.6}^{+4.0}$ & $25.9_{\;\,-8.4}^{+12.4}$\\
1.23 kpc & $1.89_{-0.42}^{+0.37}$ & $18.4_{-31.6}^{+31.6}$ & $0.30_{-0.37}^{+0.41}$ & $0.08_{-0.43}^{+0.44}$ & $21.3_{-21.3}^{\;\,+9.7}$ & $21.0_{-21.0}^{+16.9}$\\
1.47 kpc & $1.41_{-0.16}^{+0.17}$ & $77.5_{-13.1}^{+13.8}$ & $0.21_{-0.14}^{+0.13}$ & $0.25_{-0.29}^{+0.30}$ & $26.8\;\,_{-3.5}^{+3.3}$ & $24.4_{\;\,-4.8}^{\;\,+5.6}$\\
1.63 kpc & $2.12_{-0.21}^{+0.15}$ & $-55.1\;\,_{-7.2}^{+7.6}$ & $0.38_{-0.15}^{+0.15}$ & $0.25_{-0.14}^{+0.16}$ & $8.0\;\,_{-4.3}^{+3.0}$ & $8.1\;\,_{-4.4}^{+3.2}$\\
2.23 kpc & $1.73_{-0.31}^{+0.19}$ & $-33.3_{-20.9}^{+22.6}$ & $-0.31_{-0.23}^{+0.21}$ & $0.14_{-0.32}^{+0.26}$ & $8.4\;\,_{-3.9}^{+3.4}$ & $7.3\;\,_{-3.4}^{+3.4}$\\
\hline
All & $1.39_{-0.04}^{+0.04}$ & $-26.7\;\,_{-7.0}^{+6.7}$ & $0.11_{-0.15}^{+0.16}$ & $0.25_{-0.15}^{+0.15}$ & $60.3\;\,_{-1.6}^{+1.7}$ & $47.8\;\,_{-6.9}^{+9.0}$\\ 
\hline
\multicolumn{7}{p{0.1\textwidth}}{\bf Notes.}\\
\multicolumn{7}{p{0.89\textwidth}}{$^\mathrm{a}$ The aspect ratio of major and minor axes when approximating the distribution of \(\sigma_q\) and \(\sigma_u\) as an ellipse.}\\
\multicolumn{7}{p{0.89\textwidth}}{$^\mathrm{b}$ The angle between the major axis of the elliptical approximation of the \(\sigma_q\) and \(\sigma_u\) distributions and the \(q\text{-}u\) mean vector.}\\
\multicolumn{7}{p{0.89\textwidth}}{$^\mathrm{c}$ The \(\breve{\sigma}_{P\!A}\) values derived without considering the non-isotropic distributions of \(\sigma_q\) and \(\sigma_u\) using Equation (\ref{eq:norm2d}) based on \(\sigma_{q,u\perp}\) values shown in Table \ref{tab:sigma_tan}.}\\
\multicolumn{7}{p{0.89\textwidth}}{$^\mathrm{d}$ The \(\breve{\sigma}_{P\!A}\) values calculated by considering the non-nisotropic distributions of \(\sigma_q\) and \(\sigma_u\) through an elliptical approximation with non-zero skewness and accounting for the deviation from the prediction by Equation (\ref{eq:norm2d}) using numerical computation. These \(\breve{\sigma}_{P\!A}\) values are the same as those shown in Table \ref{tab:sigma_tan}.}\\
\end{tabular}
\end{center}
\end{table}

%\section{Using Chinese, Japanese, and Korean characters}

%Authors have the option to include names in Chinese, Japanese, or Korean (CJK) characters in addition to the English name.
%The names will be displayed  in parentheses after the English name.
%The way to do this in AASTeX is to  use the CJK package available at \url{https://ctan.org/pkg/cjk?lang=en}. 
%Further details on how to implement this and solutions for common problems, please go to \url{https://journals.aas.org/nonroman/}.

%日本語を使う場合、確かAASフォーマットに既に書きこまれているが、begin\{document\}の直後の
%\begin{verbatim}
%\begin{CJK*}{UTF8}{ipxm}%{gbsn}
%\end{verbatim}
%と、end\{document\}直前のend\{CJK*\}を有効にすれば良い。
%但し default の gbsn フォントは日本の漢字が結構表示出来ないので、ipxm フォントを指定する。

%Overleaf の場合、コンパイラとして pdfLaTex を選択する必要がある。


\bibliography{sagittarius}{}
\bibliographystyle{aasjournal}


\end{CJK*}
\end{document}